\documentclass[lettersize,journal]{IEEEtran}
\usepackage{amsmath,amsfonts}
\usepackage{algorithm}
\usepackage{array}
\usepackage[caption=false,font=normalsize,labelfont=sf,textfont=sf]{subfig}
\usepackage{textcomp}
\usepackage{stfloats}
\usepackage{url}
\usepackage{verbatim}
\usepackage{graphicx}
\usepackage{cite}

\usepackage{ifdraft}

\usepackage{lipsum}
\usepackage{xcolor}
\setlipsum{%
  par-before = \begingroup\color{white},
  par-after = \endgroup
}

\newif\ifusefinal
\usefinaltrue  

\colorlet{correctedcolor}{red}
\colorlet{revisedcolor}{blue}

\usepackage{booktabs}
\newcommand{\ra}[1]{\renewcommand{\arraystretch}{#1}}

\usepackage{ieeefigure}
\usepackage{listings}

\usepackage{url}
\usepackage{hyperref}
\hypersetup{
  colorlinks   = true,    
  urlcolor     = blue,    
  linkcolor    = blue,    
  citecolor    = blue 	  
}

\usepackage{algpseudocode}

\usepackage{siunitx}
\newcommand{\mysecref}[1]{Section~\ref{sec:#1}}
\newcommand{\myfigref}[1]{Fig.~\ref{fig:#1}}
\newcommand{\mytabref}[1]{Table~\ref{tab:#1}}

\newcommand*\mean[1]{\bar{#1}}
\newcommand{\nm}[1]{\qty{#1}{\nano\meter}}
\newcommand{\ghz}[1]{\qty{#1}{\giga\hertz}}
\newcommand{\thz}[1]{\qty{#1}{\tera\hertz}}

\usepackage{soul}

\newcommand{\citep}{\textbf{\color{red}[C]}}

\begin{document}

\title{Scalable Wavelength Arbitration for Microring-based DWDM Transceivers}

\author{Sunjin Choi, \IEEEmembership{Student Member,~IEEE,} and Vladimir Stojanovi\'{c}, \IEEEmembership{Fellow,~IEEE}
\thanks{Manuscript received XX, XX; revised XX, XX. This work was supported in part by task 3132.015 of the Center for Ubiquitious Connectivity (CUbiC), sponsored by Semiconductor Research Corporation (SRC) and Defence Advanced Research Projects Agency (DARPA) under the JUMP 2.0 program. \textit{(Corresponding author: Sunjin Choi.)}}%
\thanks{
This work has been submitted to the IEEE for possible publication. Copyright may be transferred without notice, after which this version may no longer be accessible.
}
\thanks{S. Choi is with the Department of Electrical Engineering and Computer Science, University of California, Berkeley, CA, 94709 USA (email: sunjin\_choi@berkeley.edu).}%
\thanks{V. Stojanovi\'{c} was with the Department of Electrical Engineering and Computer Science, University of California, Berkeley, CA, 94709 USA. He is now with Ayar Labs, Emeryville, CA 94608 USA.}}%

\markboth{Journal of Lightwave Technology,~Vol.~XX, No.~X, August~2024}%
{Sunjin Choi \MakeLowercase{\textit{et al.}}: Scalable Wavelength Arbitration for Microring-based DWDM Transceivers}

\IEEEpubid{0000--0000/00\$00.00~\copyright~2021 IEEE}

\maketitle

\begin{abstract}


\ifusefinal
	This paper introduces the concept of autonomous microring arbitration, or \textit{wavelength arbitration}, to address the challenge of multi-microring initialization in microring-based Dense-Wavelength-Division-Multiplexed (DWDM) transceivers.
This arbitration is inherently policy-driven, defining critical system characteristics such as the spectral ordering of microrings.
Furthermore, to facilitate large-scale deployment, the arbitration algorithms must operate independently of specific wavelength information and be resilient to system variability.
Addressing these complexities requires a holistic approach that encompasses the entire system, from device-level variabilities to the transceiver electrical-to-optical interface—this system-wide perspective is the focus of this paper.
To support efficient analysis, we develop a hierarchical framework incorporating an ideal, wavelength-aware arbitration model to examine arbitration failures at both the policy and algorithmic levels.
The effectiveness of this approach is demonstrated in two ways: by analyzing the robustness of each policy in relation to device variabilities, and by developing an algorithm that achieves near-perfect alignment with the ideal model, offering superior robustness compared to the traditional sequential tuning method.
The simulator code used in this paper is available at \url{https://github.com/wdmsim/wdm-simulator}.

\else
	
\fi

\end{abstract}


%
%
%
%
%
%
%
%
%
%
%

%
%
%
%
%
%
%
%
%
%

\ifusefinal
    \section{Introduction}

\label{sec:Introduction}

\IEEEPARstart{W}{ith} the overarching demand for bandwidth and the maturation of silicon photonics technology, the next phase for microring-based DWDM transceiver technology targets large-scale deployment, leveraging its high-bandwidth and low-latency characteristics \cite{wade2020teraphy, liang2022energy, levy20248, xuan2023256, wang2024silicon, chang20233d}.
However, scaling up poses significant challenges in system design, particularly with regard to mass deployment scenarios.
A cost-efficient, robust solution is essential, ideally equipped with a comprehensive analytical toolkit for large-scale deployment \cite{sun2012dsent, lee2022beyond, nedovic2023methodology}.
For microring-based DWDM transceivers, one major challenge is aligning microring resonances with laser wavelengths—a problem of \textit{initialization} \cite{mak2017multivariable}—which requires judicious arbitration of the microrings to be cost-efficient, robust and autonomous.

Microring resonance control, a common technique for addressing microring initialization \cite{sun201645, li2014silicon, de2019power, padmaraju2013wavelength, grimaldi2022self, li20203}, involves sweeping the microring resonance across the tuning range and identifying the resonance alignment.
For a single microring with a laser tone, the resonance alignment is identified by the peak in intra-cavity optical power (wavelength search) \cite{sun201645, li2014silicon, de2019power}, followed by a feedback loop that stabilize the resonance wavelength (wavelength lock) \cite{padmaraju2013wavelength, grimaldi2022self, li20203}.
However, in DWDM systems, the wavelength search would yield multiple peaks due to the multi-wavelength laser, necessitating an “informed” decision to guarantee maximum wavelength allocation \cite{hattink2020automated}; we will call this \textit{wavelength arbitration}.

\IEEEpubidadjcol

Previous work has focused on circuit-level architectures \cite{hattink2020automated, hattink2022streamlined, dong2017simultaneous, wang2022electronic, kim20234}, system-level tradeoffs \cite{sun2012dsent, georgas2011addressing, krishnamoorthy2011exploiting} and wavelength allocation algorithms \cite{wang2020energy, wang2018energy, wu2018pairing}, addressing specific aspects of arbitration.
To achieve a scalable arbitration system, we introduce a \textit{hierarchical} framework for wavelength arbitration.
In our framework, arbitration is fundamentally policy-driven, with policies defining the electrical-to-optical interface of the DWDM transceiver.
The implementing algorithm and supporting circuit-level architecture then operate according to these policies.
The primary function of arbitration, therefore, is to effectively coordinate microrings while accounting for device-level variations and system-wide constraints, ensuring both robustness and autonomy.
While this holistic view is desirable, the challenge lies in efficiently integrating these policies with scalable algorithms and architectures to maintain optimal system performance under varying conditions.

To address this, we introduce an arbitration model to decouple the analysis of policy and algorithm.
This model has two components: an ideal, wavelength-aware model and a wavelength-oblivious model.
The ideal model evaluates the arbitration policy under the assumption of wavelength-awareness, ensuring policies are effective and well-defined.
In contrast, the wavelength-oblivious model evaluates the algorithm under practical conditions, where conflict resolution and wavelength allocation are performed without specific wavelength information.
This approach allows us to evaluate arbitration robustness by analyzing failure probabilities across policies and algorithms, while also developing an algorithm that aligns closely with the ideal model, achieving scalability and robustness for high-volume deployment.

This paper is organized as follows.
We begin with an overview of the system model in \mysecref{model} and provide a formal definition of robustness metrics in \mysecref{metric}.
\mysecref{analysis} compares different arbitration policies and investigates system-level tradeoffs.
\mysecref{scheme} presents our proposed arbitration algorithm and compares it with the baseline sequential tuning scheme.
\mysecref{conclusion} concludes the paper.

    \section{Generalized Wavelength-Domain Model for Wavelength Arbitration}

\label{sec:model}

\ieeefiguredoublecolumn{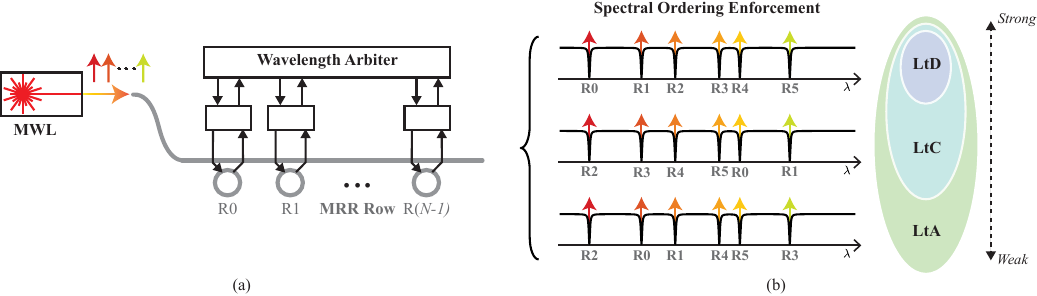}{Overview of the system model. (a) System block diagram. (b) Spectral ordering enforcement level, an arbitration policy used to classify the arbiter type. Lock-to-Deterministic (LtD) allows only a single spectral ordering. Lock-to-Cyclic (LtC) allows any cyclic equivalent of the specified ordering. Lock-to-Any (LtA) does not impose any restrictions on the final spectral ordering. These enforcement levels have an inclusive relationship, with LtA being the most permissive and LtD being the most restrictive. MWL stands for Multi-Wavelength Laser, and MRR stands for Microring Resonator.}{fig_2}

We assume a typical DWDM transceiver architecture \cite{liang2022energy} where a multi-wavelength laser propagates into a shared waveguide bus and to downstream microring resonators, with each wavelength streaming a distinct portion of data through a microring with a matching resonance wavelength.
Each microring resonance is controlled by the corresponding tuner and sensor circuits \cite{lee2022beyond, levy20248, xuan2023256}.
During transceiver initialization, each microring resonance is assigned and tuned to a unique wavelength.
This assignment process, which involves arbitrating the microring resonance wavelengths, is referred to as wavelength arbitration.
For simplicity, the process is projected onto the wavelength space; thus, only the wavelength domain is considered throughout the paper.

\subsection{Wavelength Arbiter}

\label{sec:model-arbiter}

\myfigref{fig_2}(a) illustrates the wavelength arbitration model, where the wavelength arbiter manages the global wavelength state by coordinating the wavelength states of individual microrings.
The arbiter concept encompasses two key aspects: the desired arbitration outcome and its system-level implications, referred to as the policy, and the steps required to achieve that outcome, referred to as the algorithm.
This paper focuses on a wavelength-oblivious algorithm, where the arbiter accesses local wavelength states solely through per-microring circuits.

To explain policy, we first describe the role of microring resonance wavelength in PHY-level data transmission.
Just as a port serves a transmission point in the spatial domain, a microring acts as a port in the wavelength-domain\footnote{An optical port typically refers to the fiber-coupling interface. Here, a “microring port” refers to the wavelength-domain interface that connects each DWDM channel to its corresponding electrical lane.}.
For instance, assume an electrical transmit lane drives a microring at wavelength $\lambda_{1}$, and on the receive side, another microring at resonance wavelength $\lambda_{1}$ is connected to an electrical receive lane.
When the host sends packet $\text{P}1$ through the electrical transmit lane, it is retransmitted optically at wavelength $\lambda_{1}$, and received by the receive microring, reaching the client side.
Thus, optical connectivity for $\text{P}1$ is determined by wavelength $\lambda_{1}$, serving as an optical lane for data transmission.

In the context of DWDM, microring spectral ordering can be understood as optical lane ordering (or equivalently, optical port mapping).
From a communication standpoint, transmitting and receiving endpoints should agree on the data ordering.
In microring-based DWDM systems, it is preferred that they are cognizant of optical lane ordering during electrical-to-optical and optical-to-electrical conversions\footnote{Bit ordering can be handled by MAC-level serialization and deserialization, but it is useful to track bit (lane) ordering at the transceiver level, especially for low-latency applications \cite{sharma2023system}.}.
We assume an optical port remapper (bit-shuffler backend in Georgas et al.\ \cite{georgas2011addressing}) at the optical transceiver backend to ensure the electrical interfaces at both ends see identical ordering and streamline data communication.
The level of remapping, proportional to the power and latency cost of the backend, is determined by the enforcement level over the microring spectral ordering post-arbitration.
In this paper, we adopt the enforcement level as an arbitration policy.

\subsection{Arbitration Policy}

\label{sec:model-policy}

As shown in \myfigref{fig_2}(b), we propose three wavelength arbitration policies: Lock-to-Deterministic (LtD), Lock-to-Cyclic (LtC) and Lock-to-Any (LtA), each corresponding to strong, medium and weak ordering enforcement, respectively.

\textbf{LtD policy}
demonstrates the strongest enforcement on the post-arbitration microring spectral ordering.
For example, if the target microring spectral ordering $\left(0, 1, 2, 3, 4, 5\right)$ is specified, the arbiter pairs microrings with laser wavelengths in that exact order and disallows any other spectral orderings.
By assigning microrings to a predetermined spectral ordering, it can effectively preserve optical lane ordering, bypassing any issues of optical lane shuffling.
Spectral-domain link optimization techniques can be applied more easily, such as inter-microring crosstalk minimization \cite{bahadori2016crosstalk}.
However, despite its apparent advantages, it is challenging to implement due to high microring tuning range requirements, as we will show in \mysecref{analysis}.

\textbf{LtC policy}
relaxes the enforcement by allowing cyclic equivalents of the specified target spectral ordering.
For example, if the target microring spectral ordering $\left(0, 1, 2, 3, 4, 5\right)$ is specified as in \myfigref{fig_2}, the arbiter can assign microring wavelengths in cyclic equivalents of the ordering, such as $\left(2, 3, 4, 5, 0, 1\right)$, $\left(0, 1, 2, 3, 4, 5\right)$.
The required microring tuning range is smaller than that of the LtD policy.
This is because a linear shift in spectral ordering effectively cancels any global offsets in laser and microring wavelengths due to ring resonance periodicity\footnote{Assume R0, R1, R2, R3 with \qty{4}{\nano\metre} FSR are assigned to \qtylist{1300;1301;1302;1303}{\nano\metre} laser wavelengths in the $\left(0,1,2,3\right)$ order. Due to resonance periodicity by FSR, microrings assigned to \qtylist{1301;1302;1303;1304}{\nano\metre} (shift-by-1 order) effectively shift wavelengths by \qty{1}{\nano\metre}.} \cite{georgas2011addressing, krishnamoorthy2011exploiting}.
Optical port remapping can also be done with smaller overhead using a configurable barrel-shifter in hardware \cite{georgas2011addressing} or byte-shifting in software.

\textbf{LtA policy}
does not impose any restrictions on the final microring spectral ordering.
In other words, it allows any spectral ordering post-arbitration, such as $\left(2, 0, 1, 4, 5, 3\right)$, $\left(2, 3, 4, 5, 0, 1\right)$, or $\left(0, 1, 2, 3, 4, 5\right)$.
Intuitively, it has the least requirement on microring tuning range and is most amenable to tuning power optimization techniques \cite{wang2020energy, wu2018pairing}.
However, the resulting spectral ordering is random; optical port remapping should be done by a configurable crossbar (bit-reorder mux in Georgas et al.\ \cite{georgas2011addressing}) or byte-reordering in software, potentially incurring overhead to the system.

\subsection{Multi-Wavelength Laser and Microring Row Model}

\label{sec:model-laserring}

\ieeefiguresinglecolumn{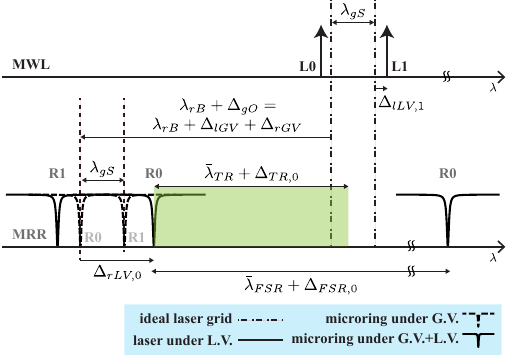}{Multi-Wavelength Laser (MWL) and Microring Resonator (MRR) row models in the wavelength domain. L.V. denotes local variation, G.V. denotes global variation, and sampled variation parameters are prefixed by $\Delta$. The tuning range (TR) and free spectral range (FSR) of the MRR are indicated in the figure. Nomenclatures are explained in \mytabref{model_params} and further detailed in the text.}{fig_3}



\begin{table}[!t]

\caption{Summary of Model Parameters}
\label{tab:model_params}
\ra{1.3}
\centering

\begin{tabular}{@{}lllp{40mm}@{}}
\toprule
\phantom{abc} & Symbol & Value & Description\\ \midrule
DWDM Grid \\
 & $N_{ch}$ & 8 & Number of DWDM channels\\
 & $\lambda_{gS}$ & \nm{1.12} & Grid spacing\\
 & $\lambda_{\text{center}}$ & \nm{1300} & Grid center wavelength\\
 & $\lambda_{rB}$ & \nm{4.48} & Microring resonance wavelength bias\\
 & $\sigma_{gO}$ & \nm{15} & Grid offset between microring row and laser grid\\
 & & & ($\sigma_{gO} = \sigma_{lGV} + \sigma_{rGV}$)\\
 \midrule
\multicolumn{3}{@{}l}{Multi-Wavelength Laser} \\
 & $\sigma_{lGV}$ & \nm{9} & Global wavelength variation\\
 & & & (Accounted in $\sigma_{gO}$)\\
 & $\sigma_{lLV}$ & 25\% & Local wavelength variation\\ 
 \midrule
\multicolumn{3}{@{}l}{Microring Resonator} \\
 & $\sigma_{rGV}$ & \nm{6} & Global resonance variation\\
 & & & (Accounted in $\sigma_{gO}$)\\
 & $\sigma_{rLV}$ & \nm{2.24} & Local resonance variation\\
 & $\mean{\lambda}_{FSR}$ & \nm{8.96} & FSR mean\\
 & $\sigma_{FSR}$ & 1\% & FSR variation\\
 & $\mean{\lambda}_{TR}$ & \textendash$^1$ & Tuning range mean\\
 & $\sigma_{TR}$ & 10\% & Tuning range variation\\
 \midrule
\multicolumn{3}{@{}l}{Microring Spectral Order} \\
 & $\left(r_{i}\right)$ & Nat.$^2$ & Pre-fabrication microring spectral ordering\\ 
 & $\left(s_{i}\right)$ & Nat. & Post-arbitration microring spectral ordering\\
\bottomrule
\multicolumn{4}{@{}l}{\footnotesize $^1$ By default, we sweep from \nm{0.28} ($0.25\times\lambda_{gS}$) to \nm{8.96} ($\lambda_{FSR}$)}\\
\multicolumn{4}{@{}l}{\footnotesize $^2$ Natural ordering i.e., $\left(0,1,2,3,4,5,6,7\right)$ for 8-Ch DWDM}\\
\end{tabular}
\end{table}

In this section, we develop a wavelength-domain model for multi-wavelength lasers and microring rows.
We first describe the pre-fabrication device model that represents the design intent of the laser and microring wavelengths.
We then present the post-fabrication model by incorporating statistical variations of the corresponding device, which we will use throughout the paper.
The model is summarized in \myfigref{fig_3} and \mytabref{model_params}.

The pre-fabrication laser model is derived from the existing commercial specifications such as CW-WDM MSA \cite{cw-wdm-msa-2023}.
According to the specification, the laser wavelengths $\lambda^{0}_{\text{laser},i}$ are uniformly spaced around a grid center wavelength, $\lambda_{\text{center}}$, with grid spacing denoted as $\lambda_{gS}$: \begin{equation} \lambda^{0}_{\text{laser},i} = \lambda_{\text{center}} + \left(i - \frac{N_{ch} - 1}{2}\right) \times \lambda_{gS} \end{equation} where $i$ is the index of the wavelength in the wavelength-domain ordering, ranging from $0$ to $N_{ch}-1$.
Default parameters for $\lambda_{\text{center}}$, $N_{ch}$ and $\lambda_{gS}$ are chosen as \nm{1300}, 8 and \nm{1.12} (\ghz{200} in O-band), respectively.
Note that a specific choice of $\lambda_{\text{center}}$ is not relevant to our focus, as only the relative distances between wavelengths matter for wavelength arbitration.

The pre-fabrication microring model is formulated similarly, with resonance wavelengths uniformly spaced by the same grid spacing, $\lambda_{gS}$, around the center wavelength, $\lambda_{\text{center}}$.
One difference is that microring wavelengths are pre-biased to the blue-side by a resonance wavelength bias ($\lambda_{rB}$) to account for the strict red-shifting of microring thermal tuning \cite{georgas2011addressing, liang2022energy}.
Additionally, for microrings, $i$ corresponds to the spatial location: the $i$th microring is the $i$th closest in the row from the light input side (R$i$ in \myfigref{fig_2}(a)).
The pre-fabrication microring spectral ordering, denoted as $r_{i}$, represents the wavelength-domain ordering of the $i$th microring at pre-fabrication.
The value of $r_{i}$ is determined by the designer’s choice for resonance wavelength placement, allowing it to deviate from the natural ordering $\left(0, 1, 2, \cdots, N_{ch}-1\right)$ for reasons such as crosstalk optimization \cite{bahadori2016crosstalk}.
Mathematically, $r_{i}$ is expressed as a list of integers, where the $i$th element corresponds to the spectral order of the $i$th microring pre-fabrication.
By default, we use the natural ordering for $r_{i}$, but also experiment with the permuted ordering $\left(0, N_{ch}/2, 1, N_{ch}/2+1, \cdots\right)$.
The pre-fabrication microring wavelength $\lambda^{0}_{\text{ring},i}$ is given by \begin{equation} \lambda^{0}_{\text{ring},i} = \lambda_{\text{center}} - \lambda_{rB} + \left(r_{i} - \frac{N_{ch} - 1}{2}\right) \times \lambda_{gS}. \end{equation}

The variation model is crucial for modeling arbitration due to the high sensitivity of photonic devices to fabrication and environmental variations.
Various models have been proposed for integrated photonics \cite{boning2022variation, lu2017performance, wang2020characterization, sun2019statistical, waqas2017sensitivity, mirza2021silicon}, but we adopt a simplified approach that considers only global (all-channel) and local (per-channel) variations.
In our model, global variations for the laser and microring are denoted as $\sigma_{lGV}$ and $\sigma_{rGV}$, respectively, while local variations are denoted as $\sigma_{lLV}$ and $\sigma_{rLV}$.
Sampled values of these variations are represented as  $\Delta_{lGV}$, $\Delta_{rGV}$, $\Delta_{lLV}$, and $\Delta_{rLV}$.

We model the variations as uniform distributions with $\sigma_{*}$ representing the half-range of the distribution.
This approach, while departing from the commonly assumed Gaussian distribution, serves as a conservative approximation of a trimmed Gaussian where the extremities are sampled more frequently.
It facilitates sample-efficient exploration of statistical bounds to achieve perfect arbitration success, and can be adjusted for use cases such as yield analysis or global sensitivity analysis.

To simplify further, we combine the global variations ($\sigma_{lGV}$ and $\sigma_{rGV}$) into a “grid offset variation”, $\sigma_{gO}$, representing the relative offset between the microring row and laser wavelengths.
This is possible because only relative distances matter in our discussion.
Since the global variations are uncorrelated, $\sigma_{gO}$ is derived as the sum of $\sigma_{lGV}$ and $\sigma_{lLV}$\footnote{$\sigma$ typically denotes statistical variance, where uncorrelated sums are sum-of-squares. For simplicity, we use a linear sum in this paper.}.
Without loss of generality, we add the grid offset to the post-fabrication laser wavelength model.

The post-fabrication multi-wavelength laser $\lambda_{\text{laser},i}$ can be written as \begin{equation} \lambda_{\text{laser},i} = \lambda_{\text{center}} + \left(i - \frac{N_{ch} - 1}{2}\right) \times \lambda_{gS} + \Delta_{gO} + \Delta_{lLV,i} \end{equation} where $\Delta_{gO}$ denotes the sampled grid offset and $\Delta_{lLV,i}$ denotes the sampled local variation of the $i$th laser wavelength.
The default parameter for $\sigma_{lLV}$ corresponds to the channel bandwidth in the CW-WDM MSA specification \cite{cw-wdm-msa-2023}, which is 25\% of the grid spacing ($\lambda_{gS}$).
This value is a conservative estimate since reported multi-wavelength laser sources have consistently achieved lower local variations \cite{sysak2022uncooled, huang20218, wang202416, zhao2024multi}.
$\sigma_{lGV}$ can be understood as the sum of center wavelength offset and variation range in CW-WDM MSA, which corresponds to the fabrication and environmental variations and specified as \nm{5} and \nm{4}, respectively.
Thus, we assume the total laser global variation ($\sigma_{lGV}$) to be \nm{9}.
For $\sigma_{rGV}$, we choose a default value of \nm{6}, which we explain in the next paragraph.

The post-fabrication microring wavelength $\lambda_{\text{ring},i}$ is expressed as \begin{equation} \lambda_{\text{ring},i} = \lambda_{\text{center}} - \lambda_{rB} + \left(r_{i} - \frac{N_{ch} - 1}{2}\right) \times \lambda_{gS} + \Delta_{rLV,i} \end{equation} where $\Delta_{rLV,i}$ denotes the sampled local variation of the $i$th microring wavelength.
It is useful to note that most of the microring variations arise from global resonance variation (wafer-to-wafer and inter-die \cite{wang2020characterization}), which can be partially mitigated by barrel-shifting the microring spectral ordering.
Global process variations can be as large as \nm{12} \cite{sun2019statistical} due to substrate thickness and etch depth variations during fabrication \cite{sun2019statistical, selvaraja2009subnanometer, mirza2021silicon, mirza2024experimental}.
However, we assume the total microring global variation ($\sigma_{rGV}$) to be \nm{6}, since process variation can also be mitigated by die binning \cite{sun2019statistical}.
External thermal aggressions can also impact global resonance variation; however, it is difficult to characterize this effect without a well-defined deployment environment, such as that described by Lee et al.\ \cite{lee2022beyond}, which we do not include in our analysis.
Microring local variation ($\sigma_{rLV}$), often referred to as channel-to-channel spacing variation, results from intra-die process variation and systematic error.
Systematic error refers to any deviation from ideal channel spacing that arises during the implementation of resonance stepping, which is typically achieved by incrementing ring radii.
This process is particularly prone to design-time errors.
In addition, intra-die process variation has a strong dependency on the physical distance between the microrings \cite{chrostowski2014impact, lu2017performance, sun2019statistical, mirza2021silicon}, making it difficult to generalize across designs.
As a result, it is challenging to extract a simple model for local resonance variation.
Instead of pursuing a detailed local variation model, we employ a zeroth-order approach, selecting a single-valued $\sigma_{rLV}$ as an upper bound to capture worst-case variation.
The default value is set to \nm{2.24} which is $2\times\lambda_{gS}$.
However, for most cases, we sweep this value from \nm{0.28} ($0.25\times\lambda_{gS}$) to \nm{8.96} ($8\times\lambda_{gS}$) to observe the system’s behavior under different microring variations.
The default parameter for $\lambda_{rB}$ is chosen as \nm{4.48} to keep the microring resonances lower than the laser wavelengths \cite{georgas2011addressing}; however, if variations exceed the fabrication bias, resonances from the preceding FSR can be utilized to be tuned to the target laser wavelength.

The tuning range of the $i$th microring can be represented as a union of intervals with a width corresponding to the tuning range and spaced by the FSR, the wavelength-domain resonance periodicity: \begin{align} \Lambda_{TR,i} &= \bigcup_{j} \left[\lambda_{\text{ring},i} + j \cdot \lambda_{FSR,i}, \lambda_{\text{ring},i} + j \cdot \lambda_{FSR,i} +\lambda_{TR,i} \right] \end{align} where $\Lambda_{TR,i}$ denotes the total tuning range, $\lambda_{FSR,i}$ denotes the FSR of the $i$th microring, $\lambda_{TR,i}$ denotes the tuning range of the $i$th microring, and $j \in \left[0,\pm1,\ldots\right]$.
Additionally, $\lambda_{FSR,i}=\mean{\lambda}_{FSR} + \Delta_{FSR,i}$ and $\lambda_{TR,i}=\mean{\lambda}_{TR} + \Delta_{TR,i}$, where $\Delta_{FSR,i}$ is the sampled FSR variation at the $i$th microring and $\Delta_{TR,i}$ is the sampled tuning range variation at the $i$th microring.
In a typical DWDM architecture, microring FSR is maximally filled with wavelength grids, or $\mean{\lambda}_{FSR} = N_{ch} \times \lambda_{gS}$, which is the value we choose (\nm{8.96} or \thz{1.6} in O-band).
The variation in the FSR ($\sigma_{FSR}$) is known to be relatively well-controlled, typically around 1\% \cite{krishnamoorthy2011exploiting, mirza2024experimental, james2023scaling}.
By default, we sweep the tuning range ($\mean{\lambda}_{TR}$) from \nm{1.12} ($1\times\lambda_{gS}$) to \nm{10.08} ($9\times\lambda_{gS}$) \cite{coenen2022thermal} and observe the trend.
For tuning range variation ($\sigma_{TR}$), we assume a 10\% variation based on the typical process, voltage and temperature variations of the tuner circuit.

The post-arbitration target spectral ordering, denoted by $s_{i}$, specifies the desired wavelength-domain ordering of the $i$th microring after arbitration.
Like the pre-fabrication spectral ordering $r_{i}$, $s_{i}$ is a list of integers, where the $i$th element corresponds to the target ordering of the $i$th microring.
As discussed in \mysecref{model-policy}, the level of enforcement on the final spectral ordering depends on the policy: LtD mandates a fixed ordering, LtC permits cyclic permutations, and LtA allows any ordering.
Mathematically, this corresponds to the $i$th microring being assigned the $j$th laser tone such that $\lambda_{\text{laser},j} \in \Lambda_{TR,i}$.
The list of $j$ forms the final spectral ordering and must equal $s_{i}$ for LtD, or be cyclically equivalent to $s_{i}$ for LtC.
By default, we assume $s_{i}=r_{i}$ under LtD or LtC, assuming the designer aligns the pre-fabrication order with the intended target ordering.

    \section{Evaluation Methodology}

\label{sec:metric}

In this section, we introduce the evaluation methodology for arbitration robustness.
Previously, tuning strategies have been evaluated based on their resulting tuning power, which facilitates comparisons of algorithm \textit{efficiency} \cite{georgas2011addressing, wang2020energy, wang2018energy, wu2018pairing}.
However, tuning power assumes arbitration success, whereas evaluating robustness requires accounting for arbitration failures.
Moreover, policy and algorithm represent distinct aspects of arbitration design—system specification and implementation—and their separation must be accounted for in an efficient, hierarchical analysis.
This necessitates a more detailed approach to devising simulation experiments and assessing arbitration \textit{robustness}.

\subsection{Policy Evaluation}

\label{sec:metric-afp}

\ieeefiguresinglecolumn{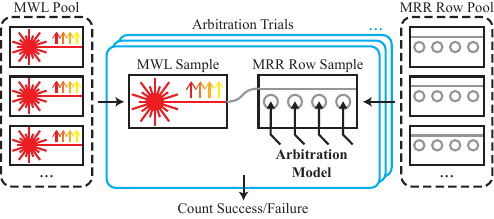}{Simulation setup for measuring the robustness of wavelength arbitration through two metrics: Arbitration Failure Probability (AFP) and Conditional Arbitration Failure Probability (CAFP). The experiments sample multi-wavelength lasers (MWL) and microring resonator (MRR) rows and subject them to arbitration tests under different policies. Policy-level evaluation uses the ideal arbitration model, which assumes wavelength-awareness and calculates AFP based on failure statistics. Algorithm-level evaluation, on the other hand, employs a wavelength-oblivious arbitration model to reflect operational constraints, computing CAFP to assess how closely the algorithm approximates ideal arbitration success.}{fig_4}

As the policy dictates wavelength arrangement post-arbitration, our policy evaluation setup assumes wavelength-awareness during arbitration, which we refer to as the ideal arbitration model.
This model arbitrates based on absolute wavelength values, effectively evaluating the arbiter’s specification independently from the wavelength-oblivious algorithm implementation.
In particular, different policies correspond to varying levels of spectral ordering enforcement, leading to distinct sets of device parameters for policy-level success.
This can be deduced from the wavelength-aware arbitration experiments with a corresponding metric defining policy-level robustness.

We define Arbitration Failure Probability (AFP) as a metric to assess arbitration robustness under specific policy and varying system configurations.
As shown in \myfigref{fig_4}, the system-under-test is generated by sampling multi-wavelength lasers and microring rows, with variations determined by their respective sources.
Each system-under-test is subjected to an arbitration test using a specific policy and the ideal arbitration model.
The number of arbitration failures is recorded, and AFP is calculated by dividing this number by the total number of trials.
Essentially, AFP reflects the arbitration yield, where failure to arbitrate successfully is treated as transceiver failure.

\subsection{Algorithm Evaluation}

\label{sec:metric-cafp}

The algorithm operates in a wavelength-oblivious manner, meaning it does not rely on specific wavelength information for arbitration.
To reflect this constraint, we employ the wavelength-oblivious arbitration model in our algorithm evaluation setup.
While AFP can be used to evaluate algorithms, it conflates the effectiveness of the algorithm with the influence of the underlying policy.
To isolate the effect of the algorithm implementation, we introduce a new metric: Conditional Arbitration Failure Probability (CAFP).

CAFP is defined as the likelihood of arbitration failure at the algorithmic level when the ideal wavelength-aware arbitration model of the corresponding policy succeeds.
By focusing on conditional algorithmic failures, CAFP isolates the algorithm’s contribution to arbitration performance and quantifies how closely the algorithm approximates the behavior of the ideal arbitration model.
To calculate CAFP, the system-under-test is generated in the same manner as for AFP evaluation.
The system is subjected to arbitration tests using both the wavelength-oblivious arbitration model (for the algorithm) and the ideal arbitration model (for the policy).
Failures are recorded when the algorithm fails at policy-level success, and CAFP is computed by dividing the number of such failures by the total number of trials.

Mathematically, CAFP can be expressed as: \begin{equation} \text{CAFP} =  P_{\text{alg}|\text{succ}}(\text{fail}) P(\text{succ}) \end{equation} where $P_{\text{alg}|\text{succ}}(\text{fail})$ represents the conditional probability of algorithmic failure given ideal arbitration success, and $P(\text{succ})$ is the probability of policy-level success.
Unlike direct conditional probability $P_{\text{alg}|\text{succ}}(\text{fail})$, CAFP achieves greater sampling stability by using the total number of trials as its denominator.
This approach avoids the variability that arises when the denominator depends on the number of successful policy-level arbitrations, particularly when $P(\text{succ})$ is low.
The total failure probability, encompassing both the policy and the algorithm, can be expressed as the sum of AFP and CAFP: \begin{IEEEeqnarray}{rCl} P_{\text{alg}}(\text{fail}) &=& P_{\text{alg}|\text{succ}}(\text{fail}) P(\text{succ}) + P_{\text{alg}|\text{fail}}(\text{fail}) P(\text{fail}) \nonumber\\ &=& \underbrace{P_{\text{alg}|\text{succ}}(\text{fail}) P(\text{succ})}_{\text{CAFP}} + \underbrace{P(\text{fail})}_{\text{AFP}} \end{IEEEeqnarray} where $P_{\text{alg}}(\text{fail})$ denotes the total failure probability.
Note that $P_{\text{alg}|\text{fail}}(\text{fail}) = 1$, as the algorithm will always fail when the ideal arbitration fails.

    \section{Analysis of Arbitration Policy}

\label{sec:analysis}

In this section, we compare different arbitration policies and analyze their tradeoffs.
We assume three key parameters: arbitration policy, pre-fabrication microring spectral ordering $\left(r_{i}\right)$ and target post-arbitration microring spectral ordering $\left(s_{i}\right)$.
We select four main test cases for policy evaluation, namely LtA-N/A, LtA-P/A, LtC-N/N, and LtC-P/P, which are summarized in \mytabref{arb_cfg}.
LtA and LtC stand for Lock-to-Any and Lock-to-Cyclic policies, as defined in \mysecref{model-arbiter}, with the corresponding levels of spectral ordering enforcement.
For pre-fabrication spectral ordering, we consider two specific instances: Natural and Permuted ordering.
Natural refers to a straightforward ordering of $\left(0, 1, 2, \cdots, N_{ch}-1\right)$ for $N_{ch}$-DWDM.
Permuted refers to $\left(0, N_{ch}/2, 1, N_{ch}/2+1, \cdots\right)$, representing a sufficiently “shuffled” ordering instance.
For post-arbitration spectral ordering, we assume the designer aims to match the pre-fabrication ordering\footnote{A case where post-arbitration spectral ordering deviates from the pre-fabrication ordering, either non-equivalently or non-cyclically ($s_{i} \neq r_{i}$), is beyond the scope of this paper, such as “channel reconfiguration” \cite{kim20234}.}, except in the case of the LtA policy, where post-arbitration spectral ordering is labeled “Any” indicating no specific enforcement.
For each experiment, we conduct 10,000 trials, using 100 multi-wavelength lasers and 100 microring row samples.
Unless otherwise noted, the default model parameters listed in \mytabref{model_params} are used throughout the experiments.


\begin{table}[h]

\caption{Arbitration Test Parameters}
\label{tab:arb_cfg}
\ra{1.3}
\centering

\begin{tabular}{@{}lllll@{}}
\toprule
\textbf{Configuration} & LtA-N/A & LtA-P/A & LtC-N/N & LtC-P/P\\ 
\midrule
 Policy & \multicolumn{2}{c}{Lock-to-Any} & \multicolumn{2}{c}{Lock-to-Cyclic} \\
 $\left(r_i\right)^1$ & Natural$^3$ & Permuted$^3$ & Natural & Permuted\\
 $\left(s_i\right)^2$ & Any$^3$ & Any & Natural & Permuted\\
\bottomrule
\multicolumn{5}{@{}l}{\footnotesize $^1$ Pre-fabrication microring spectral ordering (see \mytabref{model_params})}\\
\multicolumn{5}{@{}l}{\footnotesize $^2$ Post-arbitration microring spectral ordering (see \mytabref{model_params})}\\
\multicolumn{5}{@{}l}{\footnotesize $^3$ Labels explained in the text above}\\

\end{tabular}
\end{table}

\subsection{General Trend}

\label{sec:analysis-trend}

\ieeefiguredoublecolumn{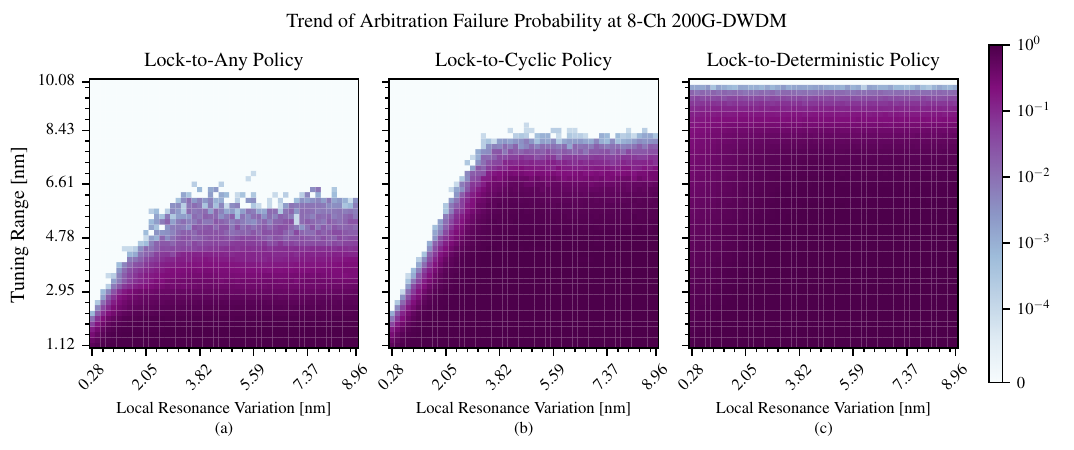}{General shmoo trend of Arbitration Failure Probability for different arbitration policies. Model parameters are shown in \mytabref{model_params}. The choice of sweep ranges for local resonance variation ($\sigma_{lLV}$) and tuning range ($\mean{\lambda}_{TR}$) are explained in \mysecref{model-laserring}.}{fig_5}

\ieeefiguredoublecolumn{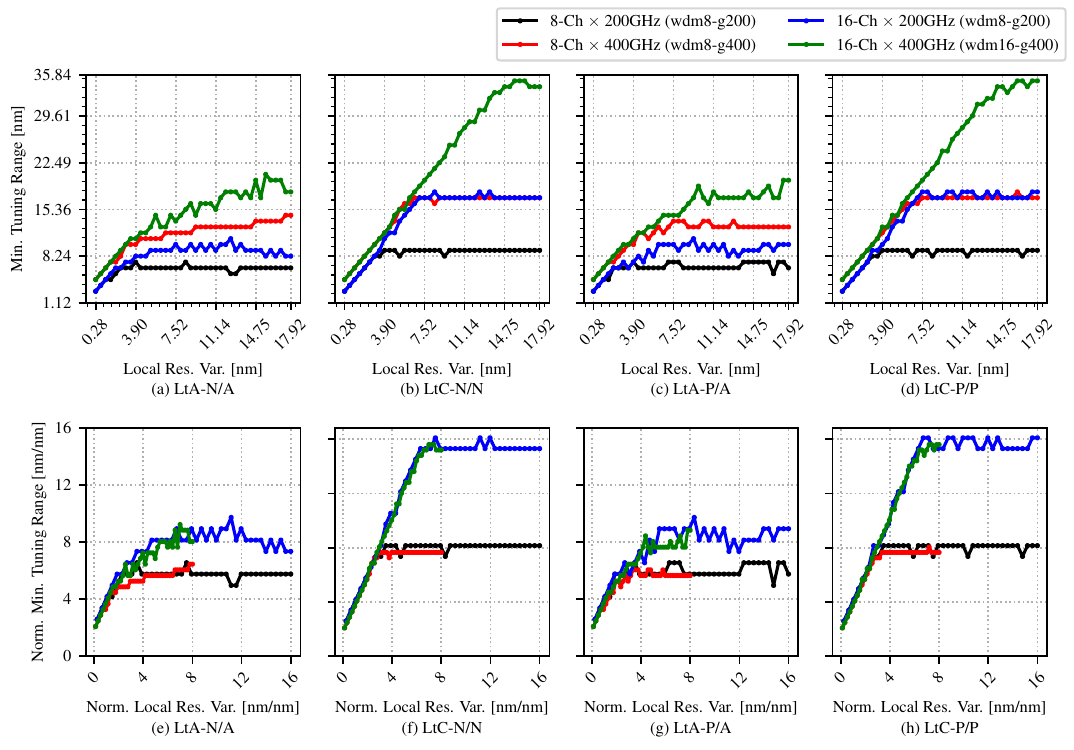}{Comparison of minimum tuning range for different DWDM and arbitration parameters. Different colors represent specific sets of DWDM parameters: channel count ($N_{ch}$) of 8 or 16 (wdm8/16) and channel spacing ($\lambda_{gS}$) of \ghz{200} or \ghz{400} (g200/400). The arbiter parameters are detailed in \mytabref{arb_cfg}. The minimum tuning range is defined as the smallest microring tuning range required for complete arbitration success. (a-d) shows the minimum tuning range trend under different arbitration policies and (e-h) plots the trends of (a-d) normalized by channel spacing.}{fig_6}

\myfigref{fig_5} illustrates the AFP across local resonance variation ($\sigma_{rLV}$) and microring tuning range ($\mean{\lambda}_{TR}$) values for different arbitration policies.
In the plot, darker colors indicate higher AFP, with dark purple ($AFP=1$) representing complete arbitration failures for all trials, while brighter colors, particularly white ($AFP=0$), indicate near or complete success.
The plot exhibits a shmoo pattern: lower tuning ranges and higher resonance variations are more likely to result in arbitration failures, whereas higher tuning ranges and lower variations increase the likelihood of arbitration success.
For each microring resonance variation value, we identify the minimum tuning range that achieves complete arbitration success, which we call the minimum tuning range.
We use this minimum tuning range as a proxy to compare different policies and system configurations below.

The LtA policy has the lowest tuning range requirement, followed by the LtC and the LtD policy, reflecting how relaxed spectral ordering reduces the microring tuning range requirement.
\myfigref{fig_6} shows the tuning range trends for different DWDM configurations, including wdm8/16, which refers to systems with 8 or 16 channels ($N_{ch}$), and g200/400, indicating channel spacings of \ghz{200} or \ghz{400} ($\lambda_{gS}$).
Both LtA and LtC policies exhibit a similar behavior, with their minimum tuning ranges saturating beyond a certain resonance variation.
For LtA, saturation occurs around $\sigma_{rLV} \sim 4\times\lambda_{gS}$ for wdm8 and $\sim 8\times\lambda_{gS}$ for wdm16, as the microring resonance can be located anywhere within the FSR once the variation range, spanning $- \sigma_{rLV}$ to $+ \sigma_{rLV}$ (totaling $2\times\sigma_{rLV}$ or the microring FSR), covers the entire FSR.
LtC saturates earlier than LtA, likely because its minimum tuning range exceeds the microring FSR before the knee point of LtA’s tuning range requirement.
Beyond the microring FSR, LtC arbitration success is guaranteed due to the resonance periodicity.

Before reaching saturation, both LtA and LtC exhibit a near-linear ramp in minimum tuning range, as shown in \myfigref{fig_6}(e-h).
The slope of the ramp is approximately 2, which we explain as follows:
$\mean{\lambda}_{minTR}$ denotes the minimum tuning range at $\sigma_{rLV}$ and $\mean{\lambda}^{0}_{minTR}$ denotes the minimum tuning range when $\sigma_{rLV}=0$.
$\Delta\lambda_{rLV,i}$ denotes the sampled local resonance variation for the $i$th microring, and assume that microrings R0 through R($N_{ch}-2$) exhibit $\Delta\lambda_{rLV,i}=-\sigma_{rLV}$ while R($N_{ch}-1$) exhibits $\Delta\lambda_{rLV,i}=\sigma_{rLV}$.
Due to the target spectral ordering allowing cyclic-equivalent arrangements, LtA and LtC can achieve the minimum tuning distance of $\mean{\lambda}^{0}_{minTR}$ when all microrings exhibit a local resonance variation of $\Delta\lambda_{rLV,i}=-\sigma_{rLV}$.
However, when only R($N_{ch}-1$) exhibits $\Delta\lambda_{rLV,i}=\sigma_{rLV}$, an additional $2\times\sigma_{rLV}$ tuning distance is required, leading to a minimum tuning range of $\mean{\lambda}_{minTR} = \mean{\lambda}^{0}_{minTR} + 2\times\sigma_{rLV}$.
LtA policy shows a slower ramp beyond $\sigma_{rLV} \sim 3\times\lambda_{gS}$, likely because it can find a matching pair that does not require additional $2\times\sigma_{rLV}$ tuning distance, especially at higher resonance variations.

Across DWDM configurations, wdm16-400g requires the most tuning range, followed by wdm8-400g, wdm16-200g, and wdm8-200g for LtA.
LtC follows a similar trend, except that it has the same tuning range requirement for wdm8-400g and wdm16-200g, and it shows a larger gap between wdm16 and wdm8 configurations compared to LtA.
This suggests that, for 16-channel cases (and likely 32-channel cases and beyond), LtA exhibits more favorable scaling than LtC in terms of tuning range requirements.
However, when local variations are kept within $3\times\lambda_{gS}$, LtC becomes a competitive candidate due to its more deterministic spectral ordering.
Denser channel spacing can also reduce the tuning range requirement, but this must be balanced against inter-microring crosstalk and multi-wavelength laser costs.

Lastly, the minimum tuning range for LtA-N/A shows no significant difference from LtA-P/A, nor for LtC-N/N compared to LtC-P/P.
This can be explained using a simple 2-WDM model:
If two arbitration trials with microrings R0 and R1 are identical but one has a pre-fabrication spectral ordering $r_{i}=\left(0,1\right)$ and the other has $r_{i}=\left(1,0\right)$, then the required tuning for R0 and R1 will effectively swap, but the minimum of two would be the same for both trials.
Thus, for ideal wavelength-aware arbitration, the difference in pre-fabrication ordering (with post-arbitration ordering $s_{i}=r_{i}$ for LtC) does not affect the minimum tuning range for LtA and LtC.
This suggests that pre-fabrication and post-arbitration spectral ordering can be flexibly assigned, and their selection is not necessarily constrained by tuning range requirements.

Next, we turn to case-studies to explore the impact of individual design parameter in greater detail.

\ieeefiguresinglecolumn{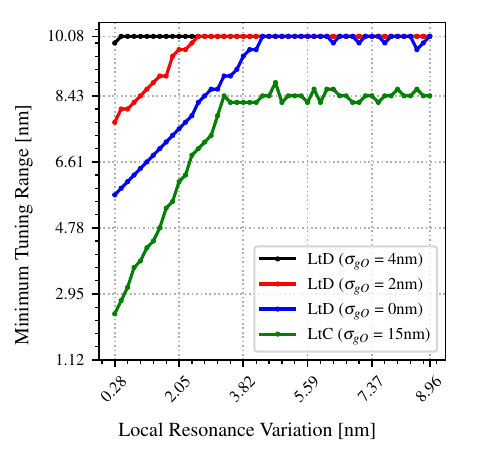}{Comparison of Lock-to-Deterministic (LtD) policy at different grid offsets ($\sigma_{gO}$).}{fig_7}

\subsection{Case of Lock-to-Deterministic Policy}

\label{sec:analysis-ltd}

The LtD policy requires a significantly larger microring tuning range than both the LtC and LtA policies to achieve perfect arbitration success, as demonstrated in previous work \cite{georgas2011addressing} and \myfigref{fig_5}.
Since LtD requires microrings to be tuned by at least the grid offset to match the corresponding laser wavelengths, we analyze the dependency of the minimum tuning range on the grid offset, as illustrated in \myfigref{fig_7}.
For small grid offsets ($\sigma_{gO}\leq$ \nm{4}), LtD exhibits a linear increase in the minimum tuning range until it saturates, whereas larger grid offsets ($\sigma_{gO}\geq$ \nm{4}) result in tuning ranges exceeding the FSR for any resonance local variation ($\sigma_{rLV}$).
The linear slope is approximately 1, because LtD requires microrings to be tuned exactly to the laser wavelength order, and when all microrings exhibit the same blue-shift ($\Delta\lambda_{rLV,i}=-\sigma_{rLV}$), the required tuning distance increases by $\sigma_{rLV}$.
As the grid offset adds directly to the required tuning distance, LtD quickly drives the minimum required tuning range beyond the FSR, which limits its practicality.
To keep the grid offset below \nm{4} and mitigate the large tuning range demands of LtD, both laser global variation ($\sigma_{lGV}$) and microring global variation ($\sigma_{rGV}$) must be controlled and minimized.
However, microring global variation alone can easily surpass \nm{4}, primarily due to substrate thickness variations (see \mysecref{model-laserring}), which may create difficulties in adopting the LtD policy.
Additionally, while the result is somewhat sensitive to the blue-shift fabrication bias of microrings ($\lambda_{rB}$ in \mytabref{model_params}), this does not significantly alter our observation.

\subsection{Impact of Laser and Microring Variabilities}

\label{sec:analysis-var}

\ieeefiguresinglecolumn{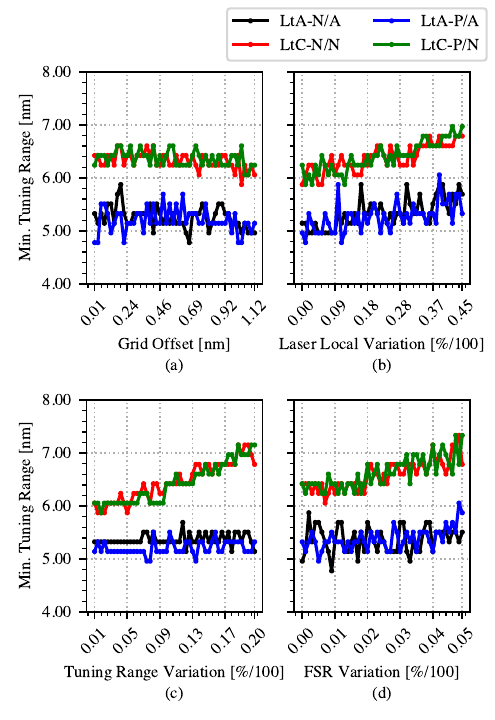}{Local sensitivity analysis on the minimum required tuning range for Lock-to-Any (LtA) and Lock-to-Cyclic (LtC) policies, considering (a) grid offset ($\sigma_{gO}$), (b) laser local variation ($\sigma_{lLV}$), (c) microring tuning range variation ($\sigma_{TR}$) and (d) microring FSR variation ($\sigma_{FSR}$). The microring local resonance variation ($\sigma_{rLV}$) is set to \nm{2.24}.}{fig_8}

\ieeefiguresinglecolumn{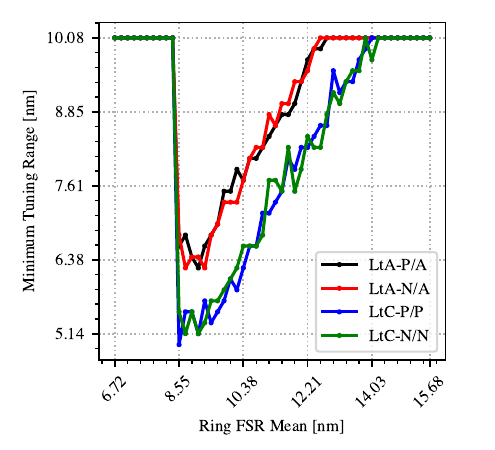}{Analysis of the impact of FSR mean ($\mean{\lambda}_{FSR}$) on minimum required tuning range.}{fig_9}

Next, we examine the impact of device variation parameters on arbitration robustness.
\myfigref{fig_8} shows the local sensitivity analysis of laser and microring variations on the minimum tuning range.
\myfigref{fig_8}(a) sweeps grid offset ($\sigma_{gO}$) from \nm{0} to \nm{1.12} (the grid spacing, $\lambda_{gS}$).
For LtA and LtC arbiters, a grid offset larger than the grid spacing does not change the required tuning distance.
This is because barrel-shifting the target spectral ordering effectively re-centers the microring row by multiples of $\lambda_{gS}$.
This adjustment cancels out that amount from the required tuning distance and compensates for grid offsets larger than $\lambda_{gS}=$ \nm{1.12}.
\myfigref{fig_8}(b) sweeps laser local variation ($\sigma_{lLV}$) from 1\% to 45\%, which nearly doubles the range of our default value, 25\%.
\myfigref{fig_8}(c) sweeps microring tuning range variation ($\sigma_{TR}$) from 0\% to 20\%, twice the nominal circuit variation range 10\%.
\myfigref{fig_8}(d) sweeps microring FSR variation ($\sigma_{FSR}$) from 0\% to 5\%, with our nominal value chosen as 1\%.

Overall, the analysis shows that the primary factors determining the required microring tuning range are microring local resonance variation and arbitration policy, while other laser and microring variations, such as tuning range variation and FSR variation, play secondary or tertiary roles.
Moreover, the trend remains consistent between pre-fabrication and post-arbitration spectral orderings, as discussed in \mysecref{analysis-trend}.
In detail, both LtA and LtC show sensitivity to laser local variation, while LtC is additionally sensitive to microring tuning range variation and FSR variation.
Denoting the minimum tuning range as $\mean{\lambda}_{minTR}$, we deduce that $\partial\left(\mean{\lambda}_{minTR}\right)/\partial\left(\sigma_{lLV}\right) \sim 0.56 \text{nm}/25\%$ for both LtC and LtA.
This can be explained by the following example:
Assume a set of laser wavelengths $\{\lambda_{\text{laser},0}, \lambda_{\text{laser},1}, \lambda_{\text{laser},2}, \lambda_{\text{laser},3}\}$ has 25\% local wavelength variations i.e., $\pm$\nm{0.28}.
Suppose $\{\lambda_{\text{laser},0}, \lambda_{\text{laser},1}, \lambda_{\text{laser},2}\}$ is shifted by $-$\nm{0.28} and $\lambda_{\text{laser},3}$ is shifted by $+$\nm{0.28}.
A microring assigned to $\lambda_{\text{laser},3}$ must shift by an additional \nm{0.56}, which roughly matches our observation.
While the sensitivities to $\sigma_{TR}$ and $\sigma_{FSR}$ in LtC arbitration can be explained similarly, it is interesting to note that in other cases, the impact of these variations is significantly reduced.
This is likely due to the looser enforcement of spectral ordering in LtA, which allows the system to “absorb” or tolerate larger variations without requiring substantial adjustments in the tuning range.

\subsection{FSR Design Guideline}

\label{sec:analysis-fsr}

Lastly, we examine the design space of microring FSR ($\mean{\lambda}_{FSR}$), as shown in \myfigref{fig_9}.
While the nominal design point is \nm{8.96}, which is $N_{ch} \times \lambda_{gS}$, $\mean{\lambda}_{FSR}$ is swept from \nm{6.72} ($6\times\lambda_{gS}$) to \nm{15.68} ($14\times\lambda_{gS}$) to investigate both under- and over-design cases.
For both LtA and LtC policies, we observe a tolerance range of approximately $\pm$\nm{0.5} around the nominal FSR, within which the increase in the minimum tuning range remains within \nm{0.5}.
For the under-designed case, exceeding this tolerance range causes a sharp increase in the required tuning range.
This is likely due to resonance aliasing, where an $\mean{\lambda}_{FSR}$ smaller than $N_{ch} \times \lambda_{gS}$ may cause a single microring to align with multiple laser wavelengths, particularly under a local laser variation of 25\%\footnote{James et al.\  \cite{james2023scaling} proposed a multi-FSR scheme that allows $\mean{\lambda}_{FSR}$ to be a fraction of $N_{ch} \times \lambda_{gS}$. However, this is beyond the scope of our analysis.}.
Over-designing the FSR results in a more gradual increase in the required tuning range.
This occurs because as the FSR widens, the distance between the last microring grid and the first microring grid of the next FSR increases, which in turn increases the required tuning distances for the LtA and LtC policies.
Therefore, to minimize AFP, it is preferable to design the FSR close to $N_{ch} \times \lambda_{gS}$, with small deviations being acceptable.

    \section{Wavelength Arbitration Algorithm}

\label{sec:scheme}

\ieeefiguredoublecolumn{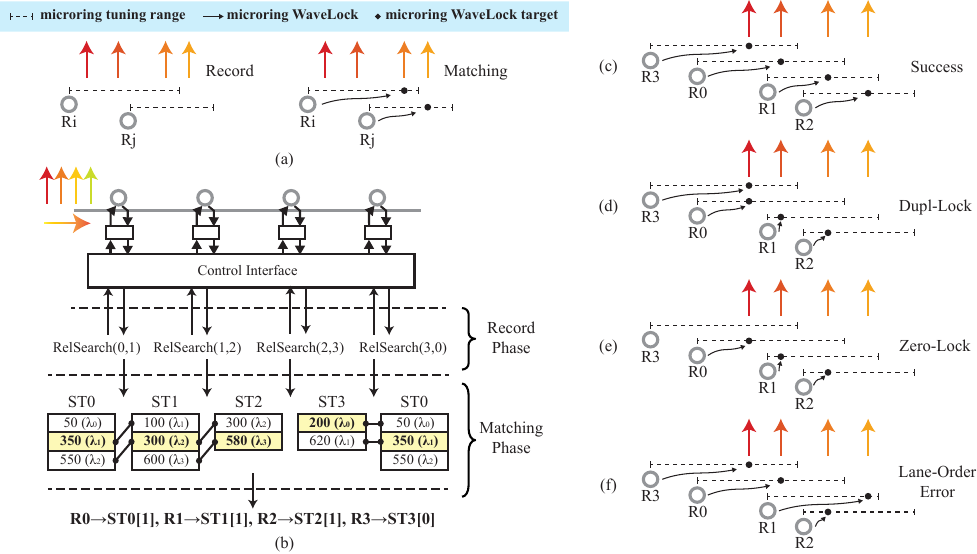}{Proposed wavelength-oblivious arbitration scheme. (a) Summary of the scheme. (b) Detailed illustration of each phase of the scheme. RelSearch stands for microring relation search. Search Table (ST) entries record the tuner codes corresponding to the peak intra-microring optical power observed during the wavelength sweep (corresponding target wavelength values are shown for illustration purposes). (c)-(f) List of arbitration outcomes.}{fig_10}

This section presents a wavelength arbitration algorithm that implements the Lock-to-Cyclic (LtC) policy\footnote{See \mysecref{scheme-discussion} for a discussion on extensions to Lock-to-Deterministic (LtD) and Lock-to-Any (LtA) policies.}.
The algorithm leverages the relative ordering of microrings and lasers in the wavelength-domain to achieve robust arbitration.
It generalizes the autonomous algorithm described in Hattink et al.\ \cite{hattink2020automated}, focusing on concepts such as target spectral ordering, conflict resolution, and maximum matching.
Throughout this section, we denote the target laser wavelengths as $\lambda_{i}$, where $\lambda_{i} < \lambda_{j}$ for $i < j$.

\subsection{Overview}

\label{sec:scheme-overview}

The algorithm consists of two phases: record and matching, as illustrated in \myfigref{fig_10}(a).
The record phase tabulates the wavelength-domain relationship between the microring wavelength search outcomes of microring pairs.
It is followed by the matching phase, where each microring is assigned to different laser wavelengths according to the identified relationships and the specified microring spectral ordering.
For convenience, we refer to the relationship between microring wavelength search outcomes as the microring relation, and the process of identifying this relationship as the relation search.
\myfigref{fig_10}(b) summarizes the structure of the scheme.

During the record phase, $N_{ch}$ (4 in \myfigref{fig_10}(b)) relation searches are conducted to discover the microring relations between consecutive microring pairs.
These pairs are derived from the target spectral ordering $s_{i}$.
For example, in \myfigref{fig_10}(b), the target pairs are $\left(0,1\right)$, $\left(1,2\right)$, $\left(2,3\right)$, and $\left(3,0\right)$ given $s_{i}=\left(0,1,2,3\right)$.
If $s_{i}=\left(0,2,1,3\right)$, the pairs are $\left(0,2\right)$, $\left(2,1\right)$, $\left(1,3\right)$ and $\left(3,0\right)$.
We note that relation searches are not run across all $N_{ch}\times(N_{ch}-1)/2$ microring pairs to construct a comprehensive “relation map” of wavelength search outcomes.
Instead, relation searches are conducted on $N_{ch}$ microring pairs.
This is sufficient because wavelength allocation for the LtC policy is \textit{semi-deterministic}; if consecutive microring pairs from $s_{i}$ are allocated to neighboring wavelengths, then the wavelengths are assigned as desired.
For example, when $s_{i}=\left(0,1,2,3\right)$, R$\left(i+1\right)$ is assigned the next laser wavelength of that assigned to R$i$.
If $s_{i}=\left(0,2,1,3\right)$, then R2 ($s_{2}=1$) is assigned next to R0, followed by R1 ($s_{1}=2$) and R3 ($s_{3}=3$).
This property allows us to simplify the global arbitration problem into $N_{ch}$ local arbitrations between consecutive pairs without compromising robustness.

The matching phase performs a non-iterative microring-to-laser matching by building a global lock allocation table.
The table is constructed by rearranging microring search tables based on the relations between consecutive pairs established during the record phase.
From this table, the final wavelength allocation is derived in a single pass, with each microring assigned its lock target from its respective search table.

Before we delve into each phase, we define the nomenclature that are used in the rest of the section.
The search table in \myfigref{fig_10}(b) represents the recorded tuner codes of the microring corresponding to the peak intra-cavity optical power during wavelength search.
We denote the search table of the $i$th microring as ST$\left(i\right)$.
A microring relation connects the two search tables and is mathematically formulated as the relation index, representing the offset between the search table entry locations according to the wavelength correspondence.
We denote the relation index between the $i$th and the $j$th microring as $RI\left(i, j\right)$.
The relation index is used to align the search tables by wavelength alignment, forming the lock allocation table for efficient arbitration.
\myfigref{fig_10}(c) depicts the arbitration success, while \myfigref{fig_10}(d-f) illustrates different failures cases: two microrings are assigned to the same wavelengths (Dupl-Lock in \myfigref{fig_10}(d)), one or more microrings are not assigned to any wavelength (Zero-Lock in \myfigref{fig_10}(e)), and lane-order mismatch or microring spectral ordering requirement is not met (Lane-Order Error in \myfigref{fig_10}(f)).

\subsection{Record Phase}

\label{sec:scheme-record}

\ieeefiguredoublecolumn{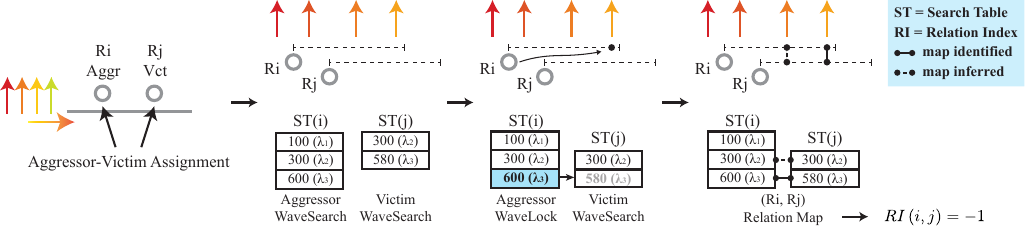}{Unit microring relation search to determine the wavelength-domain relationship between two microrings’ Search Tables (ST) through aggressor injection. The outcome of the unit relation search is a relation map between Search Tables, represented by the Relation Index (RI). A full relation search, which consists of multiple unit relation searches, is explained in detail in the text.}{fig_11}

\ieeefiguredoublecolumn{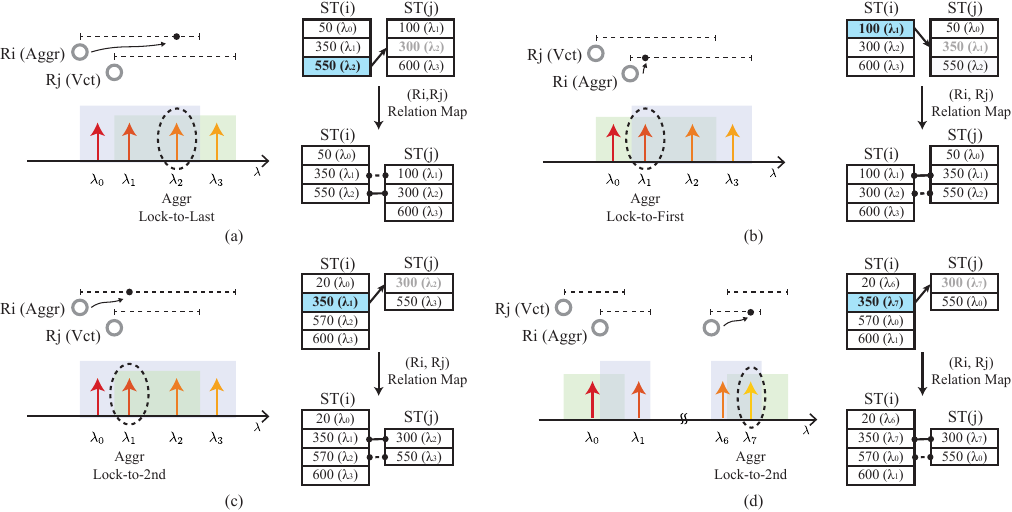}{Valid aggressor lock target during relation search depending on the relationship between the aggressor and victim microring’s resonance wavelengths ($\lambda_{aggr}$ and $\lambda_{vct}$, respectively) : (a) $\lambda_{aggr} < \lambda_{vct}$. (b) $\lambda_{aggr} > \lambda_{vct}$. (c) $\lambda_{aggr} \simeq \lambda_{vct}$ with high tuning range variation. (d) $\lambda_{aggr} \simeq \lambda_{vct}$ with high FSR variation. Blue region: aggressor microring tuning range. Green region: victim microring tuning range.}{fig_12}

As described in \mysecref{scheme-overview}, the record phase consists of $N_{ch}$ relation searches on microring pairs defined by the target spectral ordering $s_{i}$.
For the $i$th relation search, the $i$th microring (R$i$) is paired with R$j$ where $j$ corresponds to $s_{j}=s_{i}+1 \pmod{N_{ch}}$, representing an adjacent pair in the spectral ordering.

To determine the wavelength-domain relation between the search tables, the relation search identifies entries in the search tables that correspond to the same wavelength.
The search procedure, as illustrated in \myfigref{fig_11}, leverages the natural precedence order in light reception among microrings: light propagating downstream first interacts with microrings physically closer to the light input, granting them priority in capturing wavelengths.
This precedence order allows microrings with higher priority to act as aggressors during the wavelength search.
Using this property, the relation search assigns the roles of aggressor and victim to the microrings R$i$ and R$j$, respectively, based on their physical arrangement where $i < j$.

To explain the relation search process, we begin with the unit relation search, as illustrated in \myfigref{fig_11}.
Both microrings first undergo a wavelength search, recording their initial search tables, denoted as ST$\left(i\right)$ and ST$\left(j\right)$.
The aggressor microring R$i$ then wavelength locks to a specific target from its search table ST$\left(i\right)$, “injecting” aggression into the victim microring R$j$.
If the aggressor target is chosen appropriately, the victim microring finds one of the entries in its search table ST$\left(j\right)$ masked, identifying that the selected target from ST$\left(i\right)$ and the masked entry in ST$\left(j\right)$ correspond to the same wavelength.
The rest of the correspondence between the search tables can be inferred, forming a complete relation map between R$i$ and R$j$.
For example, in \myfigref{fig_11}, if the 3rd entry of ST$\left(i\right)$ and the 2nd entry of ST$\left(j\right)$ correspond to the same wavelength, then the 2nd entry of ST$\left(i\right)$ and the 1st entry of ST$\left(j\right)$ must also correspond to the same wavelength, due to the monotonicity of microring tuning in the wavelength domain.
Additionally, a search table may include tuner codes where the tuned resonances originate from different FSRs.
In such cases, the inference can naturally extend to resonances across multiple FSRs.
This is because microrings share a consistent resonance periodicity, such that the relative positions of wavelengths in search tables remain consistent across microrings and FSRs.
For instance, if ST$\left(i\right)$ and ST$\left(j\right)$ included the lowest laser wavelength ($\lambda_{0}$) by tuning from resonances at different FSRs, and the 4th entry of ST$\left(i\right)$ corresponds to that wavelength, then the 3rd entry of ST$\left(j\right)$ would also correspond to $\lambda_{0}$.
While this specific case is not depicted in \myfigref{fig_11}, it shows how shared periodicity allows inference across different FSRs.
We call these identified and inferred relations a relation map between the microrings, and define the Relation Index $RI\left(i, j\right)$ as the difference in indices between the masked entry in ST$\left(j\right)$ and the aggressor target entry in ST$\left(i\right)$.

A full relation search consists of multiple unit relation searches, where each unit relation search corresponds to a different aggressor target.
The choice of the aggressor target is critical for successful relation identification; the aggressor target wavelength must lie within the victim microring’s tuning range, which is non-trivial in a wavelength-oblivious setup.
A straightforward approach is to sequentially test each entry in ST$\left(i\right)$ as the aggressor target in multiple unit relation searches.
However, the aggressor targets can be systematically determined based on the wavelength-domain considerations, as illustrated in \myfigref{fig_12}.

\ieeefiguredoublecolumn{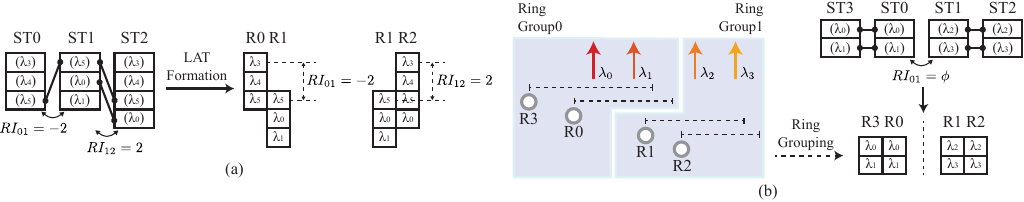}{Lock Allocation Table (LAT) formation for the matching phase. (a) Search Tables (ST) are offset by Relation Indices ($RI$) such that aligned table entries correspond to the same laser wavelengths. (b) When Relation Index is not found between the pair during the relation search ($RI=\phi$), we assume the microrings are effectively clustered and form separate Lock Allocation Tables with respect to those pairs. For brevity, only target wavelength values are shown in the Search Table and Lock Allocation Table.}{fig_13}

\ieeefiguredoublecolumn{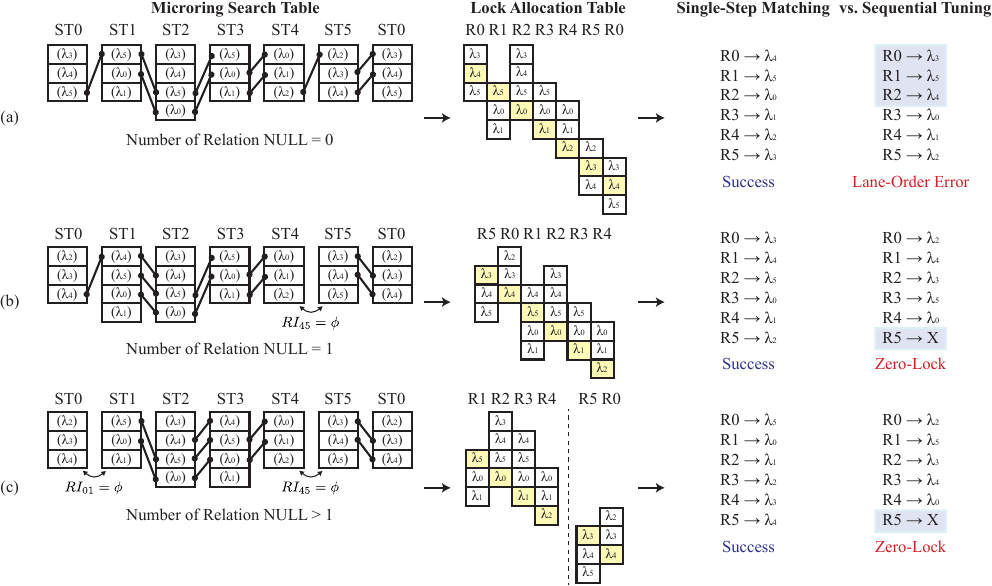}{Single-Step Matching (SSM) algorithm: (a) Zero occurrence, (b) One occurrence, and (c) More than one occurrence of Relation NULL ($RI=\phi$).}{fig_14}

\myfigref{fig_12} illustrates this selection process.
Without loss of generality, we assume $i < j$, assigning R$i$ as the aggressor and R$j$ as the victim.
In \myfigref{fig_12}(a), when $\lambda_{\text{ring},i} < \lambda_{\text{ring},j}$, setting the aggressor lock target to the last entry (Lock-to-Last) achieves the aggressor injection, as the corresponding wavelength lies within the victim’s tuning range.
Conversely, when $\lambda_{\text{ring},i} > \lambda_{\text{ring},j}$, as shown in \myfigref{fig_12}(b), setting the aggressor target to the first entry (Lock-to-First) ensures a successful relation search.
Due to process variations, the post-fabrication spectral ordering is unknown a priori.
Therefore, the Relation Search (RS) executes both unit relation searches and systematically combines their results to determine $RI\left(i, j\right)$\footnote{By combining, we mean that when both relation searches yield an $RI$ that is equivalent modulo $N_{ch}$, or when one relation search yields an $RI$ while the other fails to, the valid integer is selected as $RI$. Otherwise, the search is considered a failure.}.

In cases of large FSR or tuning range variations, neither Lock-to-First nor Lock-to-Last achieves aggressor injection.
\myfigref{fig_12}(c) illustrates the case of large tuning range variation, and \myfigref{fig_12}(d) shows the case of large FSR variation, where both the first and last aggressor targets lie outside the victim microring’s tuning range.
In these cases, we re-run the relation search with the aggressor Lock-to-Second and determine the relation.
This simple extension, which we call the Variation-Tolerant Relation Search (VT-RS), allows robust operation even under extreme variations at the cost of additional overhead.
We evaluate this in \mysecref{scheme-eval}.

\subsection{Matching Phase}

\label{sec:scheme-match}

We describe the Single-Step Matching (SSM) algorithm, a non-iterative matching algorithm designed for the LtC arbitration.
A key data structure is the lock allocation table, which provides a global view of wavelength lock candidates for microrings in the wavelength domain.

As illustrated in \myfigref{fig_13}(a), the table is constructed by organizing the search tables into a two-dimensional structure, where the horizontal axis represents different microrings and the vertical axis represents search table entries, or lock candidates.
The search tables are vertically offset by the relation index so that entries at the same vertical position correspond to the same wavelength, while the tables are arranged horizontally based on the target spectral ordering ($s_{i}$).
The algorithm then assigns lock targets to each microring through an ordered assignment, achieving a non-iterative matching that follows the specified target ordering up to cyclic equivalence.
While this approach works seamlessly when all relation searches return a valid relation index, special care is required when any relation search fails to find a relation index, which we denote as $RI=\phi$.
In such cases, we assume that the microring resonance wavelengths are “grouped” and form the sub-allocation tables according to the occurrences of $RI=\phi$.

\myfigref{fig_13}(b) illustrates the formation of sub-allocation tables and lock target assignment when certain relation searches fail to find a relation index.
In \myfigref{fig_13}(b), with $s_{i}=\left(0,1,2,3\right)$, relation searches are conducted between the pairs (R0, R1), (R1, R2), (R2, R3) and (R3, R0).
The relation searches between (R0, R1) and (R2, R3) returns empty indices ($RI\left(0, 1\right)=\phi$ and $RI\left(2, 3\right)=\phi$).
Based on these $RI=\phi$ occurrences, we interpret (R1, R2) and (R3, R0) as clustered and form sub-allocation tables accordingly.
We then assign the lock targets of R1 and R0—the aggressor/victim microrings in the relation search that yields $RI=\phi$—to the first and last entries of their respective microring search tables.
This strategy achieves the outcome of the ideal wavelength-aware LtC arbitration, which we prove using a contradiction argument:
If the wavelength-aware LtC allocation is feasible but this strategy fails, it implies the existence of an alternative valid assignment.
Without loss of generality, assume the valid assignment for R1 is an entry other than the first entry in its search table, while R0 is assigned the last entry of ST0.
R0 and R1 should be assigned neighboring wavelengths ($s_{i}=\left(0,1,2,3\right)$); However, the first entry of ST1 would lie between those wavelengths, and R0 and R1 would not be assigned neighboring wavelengths, contradicting the assumption of a valid assignment.

\myfigref{fig_14} summarizes the proposed SSM algorithm.
The algorithm is applied on a case-by-case basis, depending on the number of $RI=\phi$ occurrences, with each occurrence indicating a detected microring group.
If $RI=\phi$ does not occur, a single lock allocation table is created, and the assignment problem becomes a diagonal matching process.
When $RI=\phi$ occurs more than once, sub-allocation tables are formed, separated by the microring pairs that corresponds to $RI=\phi$.
The first microrings in each sub-table are assigned the first entries from their search tables, while the last microrings are assigned the last entries.
In case where $RI=\phi$ occurs once, a single allocation table is formed, starting and ending with the microring pair corresponding to $RI=\phi$.
\myfigref{fig_14} also compares the results of the Single-Step Matching algorithm with the sequential tuning scheme, where the first available wavelength is assigned to each microring sequentially.
Sequential tuning fails at the last microring for both \myfigref{fig_14}(b) and (c).
This failure occurs because the sequential assignment can cause earlier microrings to “steal” all available wavelengths, leaving none for the later microrings—a problem that our proposed algorithm avoids through the tabulated approach.

\subsection{Evaluation of Proposed Schemes}

\label{sec:scheme-eval}

\ieeefiguresinglecolumn{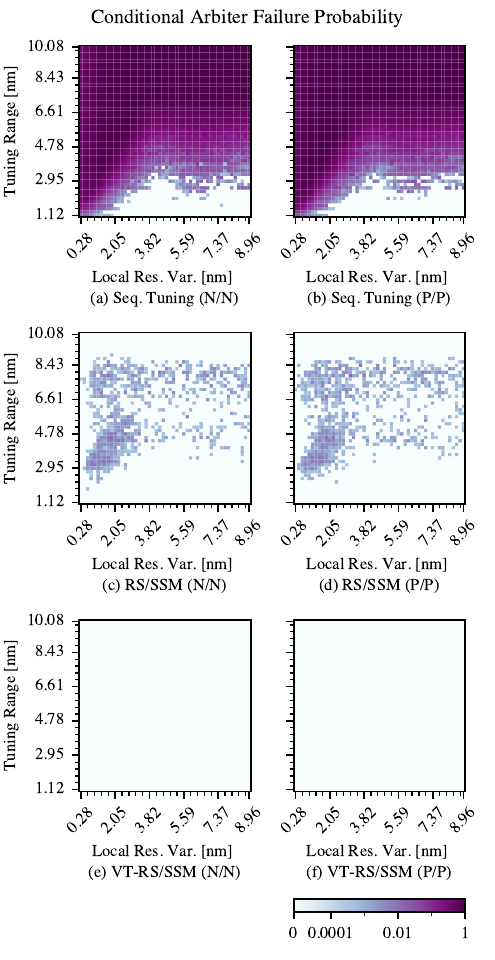}{Comparison of Conditional Arbitration Failure Probability between different arbiter schemes and configurations. Model parameters are listed in \mytabref{model_params}. Seq. Tuning refers to the sequential Lock-to-Nearest tuning scheme, which serves as the baseline, while RS/SSM (Relation Search + Single-Step Matching) and VT-RS/SSM (Variation-Tolerant Relation Search + Single-Step Matching) are the proposed methods. N/N denotes Natural ordering $\left(0,1,2,3\ldots\right)$ for both pre-fabrication spectral ordering ($r_{i}$) and post-arbitration spectral ordering ($s_{i}$). P/P denotes Permuted ordering $\left(0,4,1,5\ldots\right)$ for both $r_{i}$ and $s_{i}$.}{fig_15}

We evaluate the proposed algorithms using Conditional Arbitration Failure Probability (CAFP), as described in \mysecref{metric}.
Measured CAFP indicates how effectively the algorithm implements the LtC policy by evaluating its failure rate when the ideal LtC arbitration succeeds.
The algorithms are denoted as RS/SSM and VT-RS/SSM: RS/SSM combines the Relation Search for the record phase with the Single-Step Matching for the matching phase, while VT-RS/SSM integrates the Variation-Tolerant Relation Search for the record phase with the Single-Step Matching for the matching phase.
As in \mysecref{analysis}, our evaluation focuses on two representative target microring spectral orderings ($s_{i}$), Natural (N) and Permuted (P), and we assume that the pre-fabrication spectral ordering ($r_{i}$) matches the post-arbitration target ordering ($r_{i}=s_{i}$).
Experiments are conducted using the default model parameters listed in \mytabref{model_params}, with 10,000 arbitration tests conducted using 100 multi-wavelength laser and microring row samples.

\ieeefiguresinglecolumn{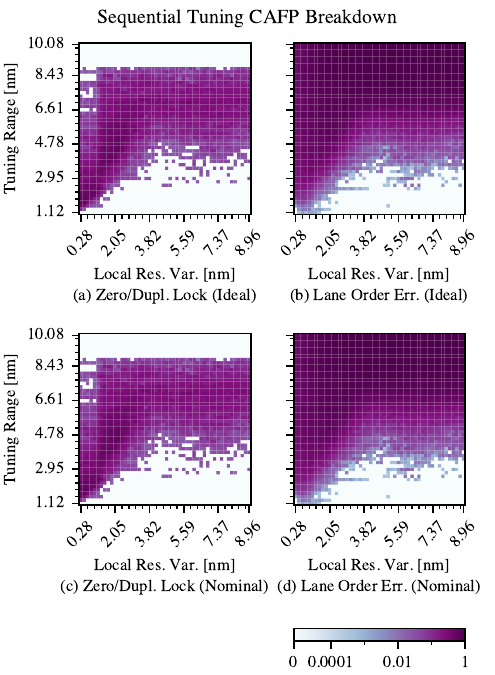}{Conditional Arbitration Failure Probability (CAFP) of sequential tuning scheme classified into Lock Error (zero-lock and duplicate-lock) and Wrong Order (lane-order error) according to \myfigref{fig_10}(d-f). (a) and (b) plots the ideal laser/microring cases where the grid offset ($\sigma_{gO}$) is set to 0 and the rest of the variations ($\sigma_{lLV}$, $\sigma_{FSR}$, $\sigma_{TR}$) are 0.1\%. (c) and (d) plots the nominal case as a comparison and used the same parameters as \myfigref{fig_15}.}{fig_16}

The baseline for our evaluation is the sequential tuning scheme, which sequentially tunes each microring to the first available wavelength at the time of assignment \cite{chang20233d, georgas2011addressing, krishnamoorthy2011exploiting, wang2023scalable}.
The order of tuning is defined by the target microring spectral ordering ($s_{i}$): after R$i$ is tuned, R$j$ is tuned such that $j$ corresponds to $s_{j}=s_{i}+1 \pmod{N_{ch}}$.
For example, for Natural ordering ($s_{i}=\left(0,1,2,3,4\ldots\right)$), R0 is tuned first, followed by R1, R2 and so forth.
For Permuted ordering ($s_{i}=\left(0,4,1,5,2\ldots\right)$), R0 is tuned first, followed by R2 ($s_{2}=1$), R4 ($s_{4}=2$) and so forth.
This baseline provides a direct comparison for evaluating how explicitly accounting for microring and laser wavelength-domain relations enhances robustness and adherence to the LtC policy in RS/SSM and VT-RS/SSM.

\myfigref{fig_15} shows the CAFP across different local resonance variations ($\sigma_{lLV}$) and microring tuning ranges ($\mean{\lambda}_{TR}$), with the choice of sweep ranges explained in \mysecref{model-laserring}.
In the plot, darker colors correspond to higher CAFP, whereas brighter colors indicate lower values.
Low CAFP is observed when $\sigma_{lLV}$ is high and $\mean{\lambda}_{TR}$ is small.
This occurs due to policy-level failures, which are not captured by CAFP.
Instead, these policy-level failures in LtC arbitration are represented by AFP, as shown in \myfigref{fig_5}(b).
Furthermore, as discussed in \mysecref{metric}, the total failure probability includes both policy-level and algorithmic failures and can be obtained by summing AFP and CAFP.

\ieeefiguresinglecolumn{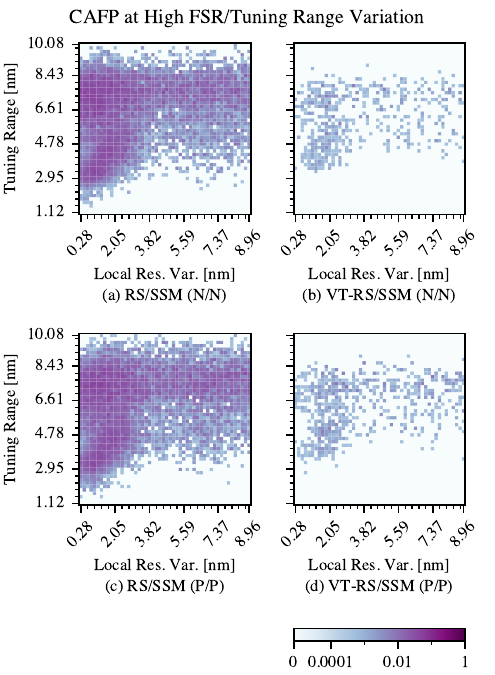}{Comparison of Conditional Arbitration Failure Probability (CAFP) for RS/SSM and VT-RS/SSM under high FSR variation ($\sigma_{FSR}$=5\%) and high tuning range variation ($\sigma_{TR}$=20\%).}{fig_17}

In \myfigref{fig_15}, we observe that the proposed schemes consistently outperform the baseline in all cases, with VT-RS/SSM closely approximating the ideal LtC arbitration.
The errors observed in RS/SSM at large tuning ranges (around \nm{8}) is likely due to a 10\% variation in tuning range, which is sufficient enough to introduce errors during the relation search, as discussed in \mysecref{scheme-record}.
Furthermore, as demonstrated in \myfigref{fig_15}(b), (d), and (f), the scheme is applicable to any target spectral ordering, assuming $s_{i}=r_{i}$.

The plotted CAFP metric includes lane-order mismatch errors (\myfigref{fig_10}(f)) due to the assumed LtC arbitration, whereas the sequential tuning scheme may not necessarily target a specific spectral ordering.
For a fair comparison, in \myfigref{fig_16}, we break down CAFP of sequential tuning into two categories: zero/duplicate lock errors (\myfigref{fig_10}(d) and (e)) and lane-order mismatch errors (\myfigref{fig_10}(f)).
When the microring tuning range exceeds approximately the FSR (\nm{8.96}), lane-order mismatch errors dominate the CAFP.
On the other hand, when the tuning range is smaller than the FSR, the sequential tuning scheme exhibits significant zero/duplicate lock errors, even under ideal laser variations and microring FSR/tuning range variations.
Empirical observations suggest that many sequential tuning failures stem from a few missing laser wavelengths.
This can be intuitively explained as follows: Suppose a natural ordering $\left(0,1,2,3,4\ldots\right)$ and due to local resonance variations, microrings R$0$, \ldots, R$\left(i-1\right)$ undergo a blue-shift, shifting their new wavelength assignments toward the previous laser tones.
If R$i$ does not experience a similar blue-shift, the assignment may skip a laser tone, resulting in a lock error.

Finally, \myfigref{fig_17} compares the CAFP of the proposed RS/SSM and VT-RS/SSM under high microring FSR and tuning range variations.
The plot suggests an interesting trend, with regions of CAFP forming around low tuning ranges (around \nm{3}) and high tuning ranges (around \nm{8}).
These regions can be linked to the relation search failure cases illustrated in \myfigref{fig_12}(c) and (d).
Specifically, we associate the region around low tuning ranges with high FSR variation, while the region around high tuning ranges would correspond to both high tuning range and FSR variations.
\myfigref{fig_17}(c) and (d) show similar regions, implying a consistent trend across different target spectral orderings.

While an FSR variation of 5\% and a tuning range variation of 20\% may be unrealistically high, \myfigref{fig_17}(b) and (d) demonstrate that VT-RS/SSM still performs well under these harsh conditions.
However, this performance comes with additional overhead, including extra steps such as aggressor Lock-to-Second and victim wavelength searches in VT-RS/SSM.
Therefore, selecting an algorithm should be done holistically, taking into account factors such as microring variations, failure tolerance levels (yield impact), and initialization costs.

\subsection{Discussion}

\label{sec:scheme-discussion}

This subsection discusses the broader implications of the proposed arbitration algorithms, including considerations for physical implementation and potential directions for future exploration.

\textbf{Physical Implementation Aspect}:
This aspect can be viewed as a practical complement to the algorithm discussions.
For instance, the proposed algorithm treats wavelength search and lock operations as unit instructions, whose implementations must consider photonics and electrical mixed-signal circuits \cite{sun201645, padmaraju2013wavelength, grimaldi2022self}.
Furthermore, after arbitration, stabilizing microring resonance is crucial to ensure modulation and demodulation quality.
This has been the focus of most prior work, including maximum Optical Modulation Amplitude wavelength lock techniques for transmitter microrings \cite{sun201645, grimaldi2022self}, and maximum power wavelength lock methods for receiver microrings \cite{lee2016errorfree}, and efficient DWDM microring wavelength lock schemes \cite{hattink2022streamlined, dong2017simultaneous, wang2022electronic}.

One of the practical challenges with the physical implementation is managing thermal crosstalk during arbitration runtime, especially in densely integrated photonic circuits.
While arbitration primarily focuses on wavelength allocation or coarse wavelength alignment—which is inherently less sensitive to thermal effects than fine alignment—thermal variability can still impact the reliability of these operations.
Addressing these challenges is crucial for robust operation at scale, and techniques such as Thermal Eigenmode Decomposition \cite{milanizadeh2019canceling} could help mitigate these effects during runtime.
However, their integration and optimization within the arbitration process remain areas for future exploration.

\textbf{Future Outlooks}:
We outline potential directions for future work.
First, the algorithm implementations of the LtD and LtA policies are left for future exploration.
For the LtD policy, a pilot signal is likely necessary to determine the absolute ordering of microring resonances.
If the algorithm relies on wavelength search outcomes, as in our proposed approach, some microrings may detect laser wavelengths from different FSRs during the search, thus limiting the ordering to modulo FSR.
For the LtA policy, we note that it has the most relaxed requirement on microring spectral ordering, providing designers with the opportunity to explore algorithms such as tuning power minimization.
Second, the framework can be extended to account for transceiver-level interactions, defining joint allocation policies for both transmitter and receiver microrings.
While this study assumes independent spectral orderings of transmitter and receiver microrings, estimating the failure rate or exploring policies with specific ordering relationships (e.g., receive-side ordering as the inverse of transmit-side ordering) represents a potential direction for future research.

    \section{Conclusion}

\label{sec:conclusion}

In this paper, we addressed the critical challenge of aligning microring resonances with laser wavelengths in microring-based DWDM transceivers, a process we termed wavelength arbitration.
We introduced an ideal arbitration model that decouples the analysis of policy and algorithm, providing a framework for evaluating the efficacy of different arbitration strategies.
Our hierarchical formulation allows for the independent assessment of arbitration policies from algorithm implementations, simplifying the analysis of both layers.

To effectively define the DWDM interface, we introduced three arbitration policies—LtA, LtC, and LtD—categorized by their spectral ordering enforcement levels and evaluated their tradeoffs in the design space.
Furthermore, we developed a wavelength-oblivious algorithm for LtC policy implementation, including the RS/SSM and VT-RS/SSM.
Through extensive simulations, we demonstrated that our schemes closely approximate the ideal model, significantly outperforming the traditional sequential tuning approach.

Our findings underscore the importance of adopting a holistic approach to wavelength arbitration, where policy-driven arbitration, alongside algorithmic implementations, enables scalable and reliable wavelength allocation.
By evaluating arbitration strategies through AFP and CAFP metrics, we provide a comprehensive view of how these strategies perform across varying device-level conditions and system configurations.
This work establishes a foundation for advanced arbitration techniques, demonstrating the need to integrate both high-level policy design and practical, wavelength-oblivious algorithm implementation.
Moving forward, further research into variations such as LtA policy implementations will be crucial for enhancing the scalability and robustness of microring-based DWDM transceivers.

\else

\fi


\section*{Acknowledgments}
This work was supported in part by task 3132.015 of the Center for Ubiquitious Connectivity (CUbiC), sponsored by Semiconductor Research Corporation (SRC) and Defence Advanced Research Projects Agency (DARPA) under the JUMP 2.0 program.
Sunjin Choi would like to acknowledge the support from Korea Foundation for Advanced Studies (KFAS) and Qualcomm Innovation Fellowship (QIF).
The authors would like to thank Shenggao Li from TSMC and Pavan Bhargava from Ayar Labs for their valuable discussions and feedback.
The authors would like to give special thanks to Yue Dai for her valuable assistance in writing the manuscript.
The authors also acknowledge the students, staff, faculty and sponsors of the Berkeley Wireless Research Center.



\bibliography{IEEEabrv,./bib/general, ./bib/singlering, ./bib/algo,./bib/impl,./bib/device}
\bibliographystyle{IEEEtran}

\newpage

 

\begin{IEEEbiography}
[{\includegraphics[width=1in,height=1.25in,clip,keepaspectratio]{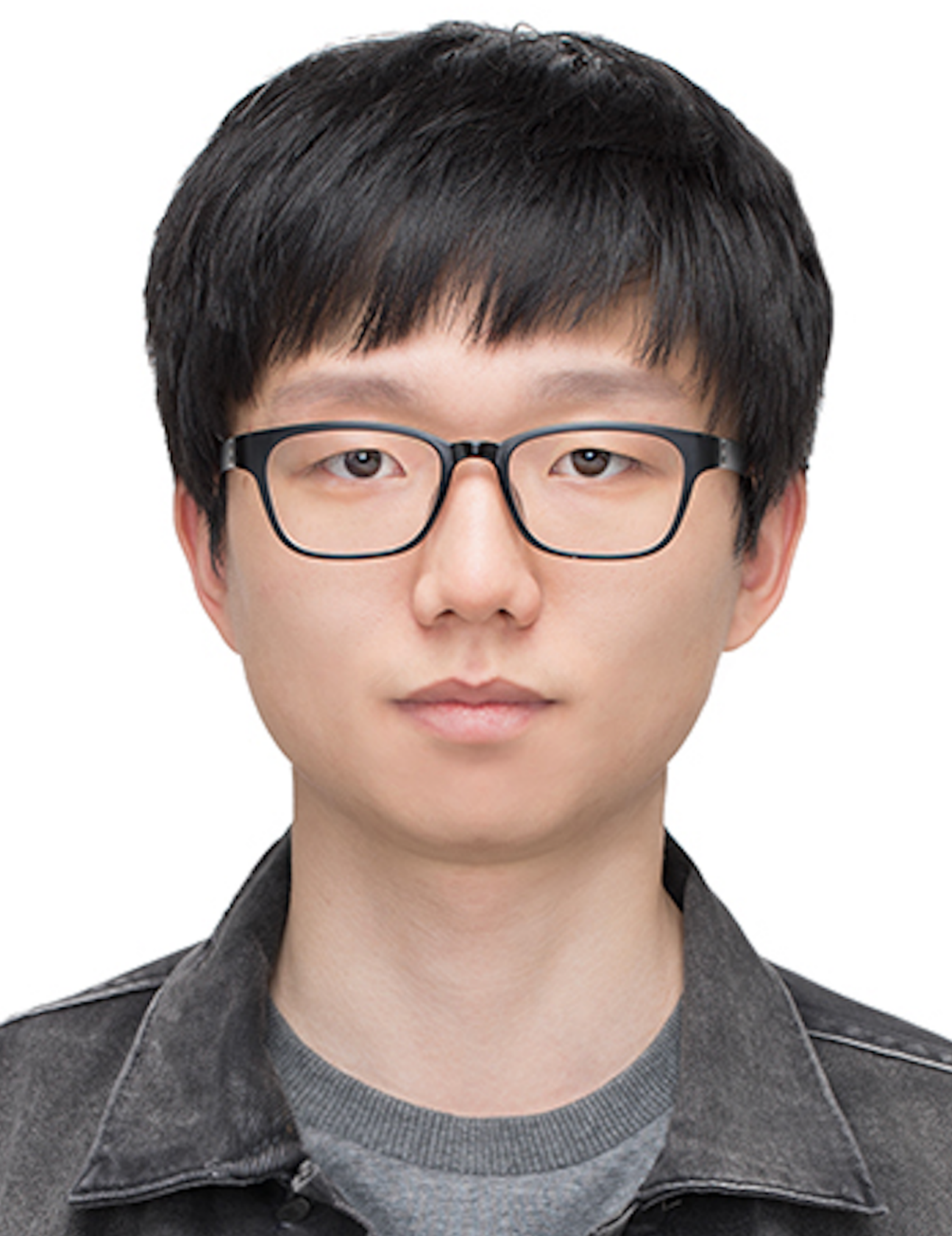}}]{Sunjin Choi}
(Student Member, IEEE) received the B.S. degree in electrical engineering and computer science from Seoul National University, Seoul, South Korea, in 2020. He is currently pursuing the Ph.D. degree in electrical engineering and computer sciences at the University of California, Berkeley, CA, USA.

He has held an internship position with Ayar Labs and X, the moonshot factory, where he worked on integrated photonic system design. His research interests include design of electronic-photonic integrated systems for high-speed optical interconnects.
\end{IEEEbiography}

\begin{IEEEbiography}
[{\includegraphics[width=1in,height=1.25in,clip,keepaspectratio]{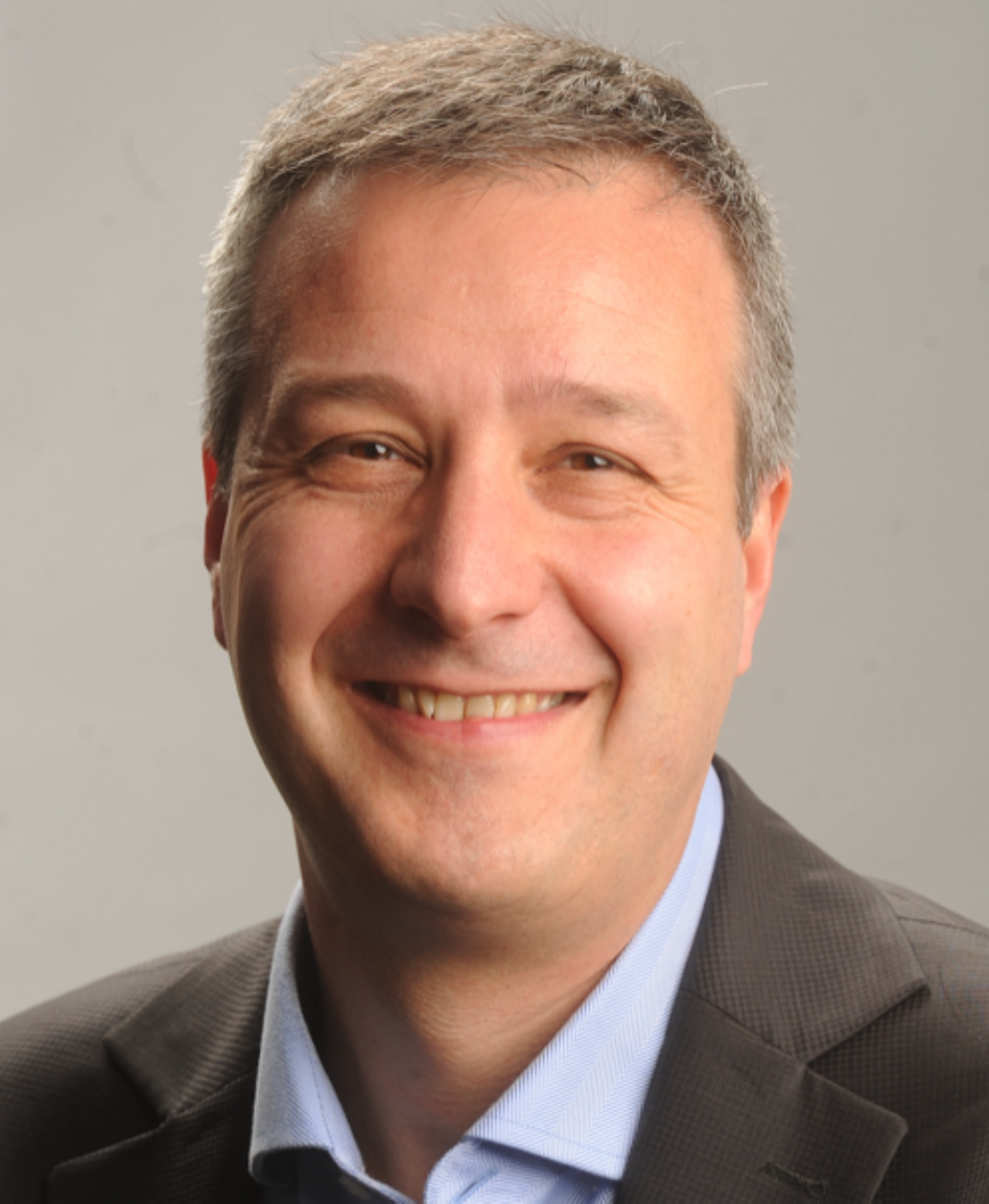}}]{Vladimir Stojanovi\'{c}}
(Fellow, IEEE) received 
the Dipl.-Ing. degree from the University of Belgrade, Belgrade, Serbia, in 1998, and the Ph.D. degree in electrical engineering from Stanford University, Stanford, CA, USA, in 2005.

He was with Rambus, Inc., Los Altos, CA, USA, from 2001 to 2004; and the Massachusetts Institute of Technology, Cambridge, MA, USA, as an Associate Professor, from 2005 to 2013. He is currently a Professor of electrical engineering and computer sciences with the University of California at Berkeley, Berkeley, CA, USA, where he is also a Faculty Co-Director of Berkeley Wireless Research Center (BWRC), Berkeley. His current research interests include the design, modeling, and optimization of integrated systems, from CMOS-based VLSI blocks and interfaces to system design with emerging devices, such as NEM relays and silicon photonics, design and implementation of energy-efficient electrical and optical networks, and digital communication techniques in high-speed interfaces and high-speed mixed-signal integrated circuit (IC) design.

Dr. Stojanovi\'{c} was a recipient of the 2006 IBM Faculty Partnership Award, the 2009 NSF CAREER Award, the 2008 ICCAD William J. McCalla, the 2008 IEEE Transactions on Advanced Packaging Award, and the 2010 ISSCC Jack Raper Best Paper and 2020 ISSCC Best Forum Presenter Awards. He was a Distinguished Lecturer of IEEE Solid-State Circuits Society from 2012 to 2013.
\end{IEEEbiography}

\vspace{11pt}

%

\vfill

\end{document}